\documentclass{aa}  
\usepackage{graphicx}
\usepackage{epsfig}
\usepackage{txfonts}
\usepackage{natbib}
\bibpunct{(}{)}{;}{a}{}{,}
\newcommand{\La}{Ly$\alpha$ }
\begin{document}
   \title{Limits on the luminosity function of Ly$\alpha$ emitters at $z = 7.7$ 
\thanks{Based on observations obtained at the Canada-France-Hawaii Telescope (CFHT) which is operated by the National Research Council (NRC) of Canada, the Institut National des Sciences de l'Univers of the Centre National de la Recherche Scientifique of France (CNRS), and the University of Hawaii. This work is based in part on observations obtained with MegaPrime/MegaCam, a joint project of CFHT and CEA/DAPNIA and in part on data products produced at TERAPIX and the Canadian Astronomy Data Centre as part of the Canada-France-Hawaii Telescope Legacy Survey, a collaborative project of NRC and CNRS.}}

   \author{
P. Hibon
          \inst{1,2}
          \and
          J.-G. Cuby\inst{1}
           \and
         J.Willis\inst{4}	
          \and
          B. Cl\'ement\inst{1}
          \and
          C. Lidman\inst{3}
          \and
          S.Arnouts\inst{5,1}
          \and
          J.-P. Kneib\inst{1}
          \and
          C. J. Willott\inst{6}
          \and
          C. Marmo\inst{7}
          \and
          H.~McCracken\inst{7}
          }

   \offprints{P. Hibon}

   \institute{Laboratoire d'Astrophysique de Marseille, OAMP, Universit\'e Aix-Marseille \& CNRS, 38 rue Fr\'ed\'eric Joliot Curie, 13388 Marseille cedex 13, France 
   				\and
Korean Institute for Advanced Study, Dongdaemun-gu, Seoul 130-722, Korea
                                \and
European Southern Observatory, Alonso de Cordova 3107, Vitacura, Casilla 19001, Santiago 19, Chile
   				\and
Department of Physics and Astronomy, University of Victoria, Elliot Building, 3800 Finnerty Road, Victoria, BC, V8P 1A1, Canada
   				\and
Canada France Hawaii Telescope Corporation, Kamuela, HI 96743, USA
   				\and
Herzberg Institute of Astrophysics, National Research Council, 5071 West Saanich Rd, Victoria, BC V9E 2E7, Canada
   				\and
Institut d'Astrophysique de Paris, Universit\'e Pierre et Marie Curie, 98bis boulevard d'Arago, 75014 Paris, France
}
   \date{}

  \abstract
   {}
{The \La luminosity function (LF) of high-redshift 
\La emitters (LAEs) is one of the 
few observables of the re-ionization epoch accessible to date with 8-10 m 
class telescopes. The evolution with redshift allows one to constrain 
the evolution of LAEs and their role in re-ionizing the Universe at the end
of the Dark Ages.}
   {We have performed a narrow-band imaging program at 1.06 $\mu$m at the CFHT, 
targeting \La emitters at redshift $z \sim$ 7.7 in the CFHT-LS D1 field. From 
these observations we have derived a photometric sample of 7 LAE 
candidates at $z \sim$ 7.7. }
   {
We derive luminosity functions for the full sample of seven objects and for 
sub-samples of four objects. If
the brightest objects in our sample are real, we infer a luminosity function
which would be difficult to reconcile with previous work at lower redshift.
More definitive conclusions will require spectroscopic confirmation.
}
   {}


   \maketitle
%
%
\section{Introduction}
Searching for high-redshift galaxies is one of the most active fields
in observational cosmology.  The most distant galaxies provide a
direct probe of the early stages of galaxy formation in addition to
revealing the effects of cosmic re-ionization  \citep{Fan2006}.
Galaxies at redshift 6 are routinely found and show that star formation was 
initiated at significantly higher redshifts and that these galaxies are the 
likely sources of the re-ionization of the universe which was completed at 
this redshift.
Conversely, detection of $z > 7$ galaxies is still rare, due in 
large part to the complete absorption of their restframe UV emission below the 
\La line which is redshifted beyond the 1 $\mu$m cutoff wavelength  of silicon.
The deployment of large format IR arrays at many telescopes now makes  these 
observations possible. From $z = 6.5$ to $z = 7.7$ light dimming due to 
luminosity distance is 30\% and the
age of the Universe varies by 150 Myr, leading to a further
dimming, probably moderate considering the relatively short time span, 
of the galaxy intrinsic luminosities.
Observations of $z > 7$ objects should therefore remain within reach of the 
current generation of telescopes.
However, at $z > 6.5$  the Universe is thought to be undergoing 
re-ionization and this may cause a further evolution of the
observable properties of very distant objects with redshift, 
possibly more abrupt than their intrinsic evolution and dimming with age
and distance.

One prime tracer of high-redshift galaxies is the \La line. The determination
of the \La Luminosity Function (LF) with infrared arrays is actively pursued by 
several groups, either through narrow-band imaging \citep[e.g][]{Willis2008, 
Cuby2007}, or through blind spectroscopy along the critical lines
of galaxy clusters used as gravitational telescopes \citep{Richard2006, Stark, Bouwens2008}. 
High-$z$ galaxies are also searched using the dropout technique 
between the optical and near infrared domains, either in the field
\citep[see e.g.][]{Bouwens} or behind galaxy clusters 
\citep[see e.g.][]{Richard}. 
The dropout method is primarily sensitive to the UV continuum emission of
the galaxies and therefore allows to determine their UV Luminosity Function
(UVLF).

The UVLF of LAEs is a direct tracer of galaxy evolution and it is not
affected by the amount of neutral hydrogen in the intergalactic medium (IGM), while
the \La emission (and therefore the \La LF) may be affected. A rapid change
in the ionization state of the Universe could lead to a decline
of the \La luminosity density at high-redshift, while the UVLF should have
a milder evolution. Evidence for such a rapid change of the
neutral fraction of the Intergalactic Medium between redshifts 6 and 7
includes the observation of LAEs in narrow-band imaging at 
$z = 6.5$ \citep{Kashikawa} and at $z = 7$ \citep{Ota} and in spectroscopy
at $z > 7$ \citep{Richard}. The  patchy structure of a partially
ionized Universe should also affect the apparent clustering of LAEs
at high redshifts, see \citet{Mesinger} for an analysis of this effect
at $z \sim 9$.

More observations of LAEs at high redshifts are necessary to better 
characterize the re-ionization epoch, and in particular observations in the
near-IR domain to probe redshifts $\ga$ 7. \citet{Willis}, \citet{Willis2008}
and \citet{Cuby2007} have performed narrow-band surveys at $z = 8.8$ 
which have yielded only upper limits of the \La LF of LAEs at this redshift.
In this paper we present the results of a narrow-band imaging survey 
at $z = 7.7$ representing a factor of 10 improvement in area at approximately 
the same detection limit compared to our previous survey at $z = 8.8$.
These observations were made with the Wide Field near-IR Camera
(WIRCam) operating at 
CFHT\footnote{See http://www.cfht.hawaii.edu/Instruments/Imaging/WIRCam/}.

In section 2, we describe the narrow-band observations and other observational 
material 
used in this paper. In section 3, we discuss the construction of our sample
of Ly$\alpha$ emitters. In section 4, we infer from this sample an estimate 
of the \La luminosity 
function of $z = 7.7$ LAEs and discuss it in the light of existing models. \\

Unless explicitly stated otherwise, we use AB magnitudes throughout the paper. 
We assume a flat $\Lambda$CDM model with $\Omega_{M} = 0.27$ and $H = 70$ 
km $\, \textnormal{s}^{-1}\, \textnormal{Mpc}^{-1}$.

\section{Observational data}
The CFHT-LS D1 field was chosen for the availability of multi-wavelength 
data from the 
X ray to the near-IR, including in particular extremely deep optical data
from the CFHT-LS survey. For the purpose of this study, we 
originally made use of the so-called
T0004 release of the CFHT-LS survey, and later  of the T0005 release when 
it became available (November 2008). The CFHT-LS data products are 
available from the CADC archive to CFHT users and consist in various
image stacks in the $u^{\ast}g'r'i'z'$ filters and of ancillary data, such as
weight maps, quality checks, catalogs, etc. The $u^{\ast}$, $g'$, $r'$, $i'$, $z'$ filters have spectral
curves similar to the SDSS filters.

The core data relevant to this paper are deep near-IR Narrow
Band (NB) observations of a $\sim 20\arcmin \times 20\arcmin$ area of the  
CFHT-LS D1 Deep Field, totalling $\sim$ 40 hrs of integration time. In addition 
to the narrow-band near-IR data we made use of  broad band $J$, $H$ and $Ks$ data 
of the field which were acquired as part of another program carried out with 
WIRCam (PIs Willott \& Kneib). We also used near-IR Spitzer/IRAC data from the 
SWIRE survey \citep{Lonsdale2003}.

\begin{figure}
\centering
\resizebox{\hsize}{!}{\includegraphics{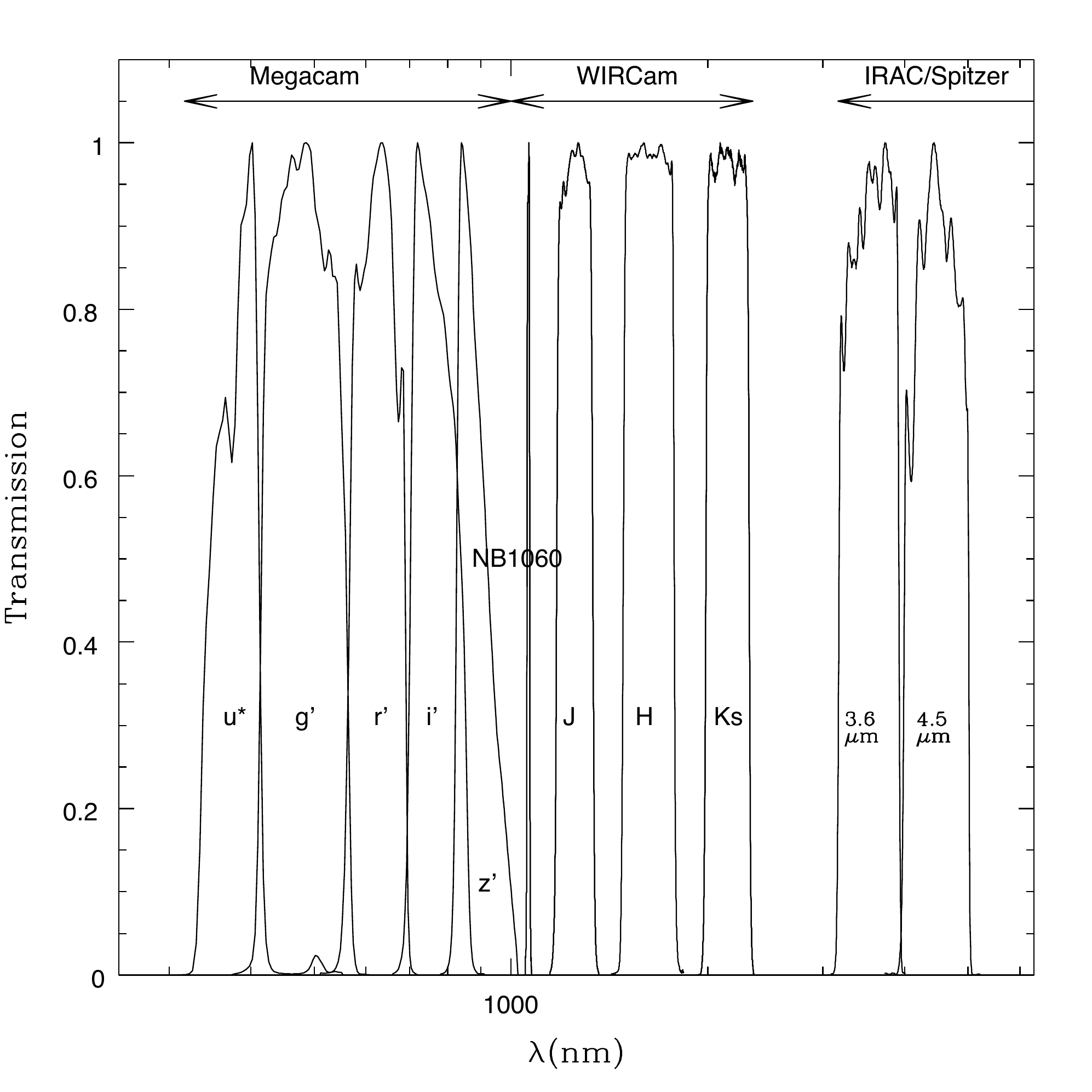}}
\caption{Transmission curves of the filters corresponding to the data
used in this paper. All transmissions are normalized to 100\% at maximum.}
\label{fig:filters}
\end{figure}

A summary of the observational data used in this paper is provided
in table~\ref{tab:mag}. Figure~\ref{fig:filters} shows the transmission curves 
of the filters corresponding to the multi-band data used in this paper.

\subsection{WIRCam narrow-band data}
WIRCam is a $\sim 20\arcmin \times 20\arcmin$ imager installed at the CFHT 
prime focus.
It is equipped with 4 Hawaii2-RG arrays, the pixel scale is 
0.3\arcsec $\, \textnormal{pixel}^{-1}$ and the four arrays are separated by a 
$\sim$ 15\arcsec\ gap. We used a Narrow Band filter centered at 1.06 $\mu$m 
(hereafter referred to as $NB1060$) with a full width at half 
maximum of 0.01 $\mu$m ($\sim$ 1\%). This filter was designed to  match a 
region of low night sky OH emission. 

The data were acquired in service mode over several months in two different 
semesters in 2005 and 2006, each epoch totalling approximately
the same  integration time ($\sim$ 20 hrs). 
A detector integration time of 630\,s was used, ensuring
background limited performance. The  sky background was measured 
to be of the order of 4\,e$^-$ s$^{-1}$ pixel$^{-1}$ which corresponds to
a sky brightness of $\sim$ 17.7 mag arcsec$^{-2}$ (Vega).
Our first observations started shortly after the commissioning of the camera.
The detector was experiencing variable readout noise and significant
electronic crosstalk, both issues that were progressively solved in the 
subsequent months of operations. The first year data were the most affected. 
We discuss in a subsequent section of this paper 
the effects of the electronic cross-talk as
a possible source of contamination.

With WIRCam the only possibility to guide the telescope is through 
the on-chip guiding capability implemented in the detector controller.
This is achieved by clocking
adequately small detector windows around bright stars for fast readout and
rapid guiding. This feature does also leave some residuals on the images along
the detector lines where the windows are located.

The narrow-band data were pre-processed at CFHT (dark subtraction and flat
fielding). The pre-processed images were then stacked together at the Terapix
data processing center at Institut d'Astrophysique de Paris (IAP). 
The data reduction steps include double pass
sky subtraction, astrometric and photometric calibration and final stacking of
the images. Two separate stacks were produced for each of the one-year datasets.
For the details of the data reduction, see \citet{Marmo2007}. We further
combined these two stacks into a single stack corresponding to the entire
dataset. After stacking, the useful area of the array after removal of the
edges was 390 arcmin$^2$. The FWHM image quality of the final stacked image,
as measured on stars, is $\sim$ 0.76\arcsec.

\subsection{WIRCam Broad band data}
The broad band $J$ and $Ks$ WIRCam data were acquired in 2006 and 2007 and were 
processed similarly to the narrow-band data. We also had moderately deep 
SOFI (ESO-NTT) $J$ and $Ks$ band images covering one quarter of the WIRCam field 
\citep{Iovino}. This allowed us to improve the limiting magnitude
by $\sim$ 0.3 mag in each of the $J$ and $Ks$ bands for the corresponding quadrant.
We did not attempt at making use of the UKIDSS 
Deep Extragalactic Survey (UKIDSS-DXS) \citep{Lawrence} 
$J$, $H$ and $Ks$ images as these
data are of significantly shallower depths. The image quality achieved in
the $J$ and $Ks$ image is comparable to that of the $NB1060$ image.

We note that the $NB1060$ and $J$ WIRCam filters do not overlap. The wavelengths
of the blue and red ends of the full width at half maximum of the $J$ filter are 1175 and 1333 nm. 
The $J$ filter can therefore be used to trace the UV continuum above the
\La line without being contaminated by line emission.

\subsection{Photometric calibration}

Due to the general lack of photometric calibration data in narrow-band filters 
and to the lack of systematic observations of standard stars with the WIRCam
$NB1060$ filter, we performed the photometric calibration of the narrow-band 
dataset using stars present in the field. For consistency we used identical 
procedures for the photometric calibration of the entire MegaCam and WIRCam 
datasets. We initially selected stars on  morphological
criteria from the list of detected objects. We then selected
those stars which were not saturated in any of our images. From this list
of objects we finally kept those present in the 2MASS catalog. This led
to a final stellar sample of 75 objects. 
We first determined the zero points of the WIRCam Broad Band data ($J$ and $Ks$)
by minimizing the difference between the WIRCam and 2MASS magnitudes of the
stellar sample. This left residuals of 0.07 and 0.15 magnitudes (rms)
in the $J$ and $Ks$ bands respectively.

For the MegaCam data, we  applied zero point offsets up to 0.06 
magnitude to the photometric
catalog distributed as part of the T0004 CFHT-LS release. These offsets
were determined by \cite{Ilbert} when trying to adjust, for the purpose of
determining photometric redshifts, the original CFHT-LS photometry to 
synthetic colors of galaxies derived from SED models. These offsets
were originally determined for the T0003 release, and we used slightly
modified ones corresponding to the T0004 release
\citep{Coupon,Ilbert}.

As a parallel investigation, we generated synthetic colors for the 
MegaCam and WIRCam filters of a variety of  stellar spectra models of various 
temperatures and metallicities 
(\citet{Marigo} and http://stev.oapd.inaf.it/~lgirardi/cgi-bin/cmd). While
the WIRCam $J$ and $Ks$ magnitudes of the stellar sample did match well to the 
synthetic color tracks, the CFHT-LS magnitudes had to be slightly modified 
by offsets very similar to those mentioned above to better match the color
tracks. This suggests that there are  systematic offsets between
the CFHT-LS photometry and synthetic colors of stars and
galaxies. That the offsets for stars and galaxies are similar is not surprising 
for synthetic SED modeling of galaxies makes direct use of
stellar spectra. More interesting is the fact that these offsets do not
seem to depend on the models being used, as noted by \citet{Ilbert}, a fact
which seems re-inforced by our analysis since we used completely different
synthetic stellar libraries.

Finally, we performed a last check using the stellar
library of Pickles \citep{Pickles}. This library consists of observed
stellar spectra in the optical domain and in 
parts of the near-IR domain, with interpolations in between.
Here again the color tracks derived for the MegaCam filters from this library 
do match the color tracks of our stellar sample after applying the same
offsets as above. 

From the color tracks we determined that the stars of the stellar 
sample used for the calibration have spectral types from G to M5.
Then, from the calibrated $u^\ast$, $g'$, $r'$, $i'$, $z'$, $J$ and $Ks$ data, 
we performed an ad hoc polynomial fitting of the fluxes of all objects
in the stellar sample, 
from which we derived for each star the $NB1060$ magnitude at 1.06 $\mu$m.
This simple method is justified
in view of the large number of photometric datapoints (7) available, of
the smooth spectral energy distribution of stars, and of
the absence of features at the wavelength of the $NB1060$ filter
in the infrared spectra of stars of  spectral types earlier than M5.

With the stellar $NB1060$ magnitudes, we could then determine the zero points 
of our $NB1060$ image. Because quadrants 
have slightly different gains, this was done separately for each quadrant.
The rms residuals after this last step were 0.04 magnitude. 

Making provision for additional sources of errors, e.g. the accuracy 
of the 2MASS photometry, possible biases from the selected sample of stars, 
etc., we estimate our final photometric accuracy to be of the order of 0.1 
magnitude rms and we adopt this value in the rest of this paper.

\subsection{Catalog generation and detection limits}
We used  SExtractor \citep{Bertin} in single image mode for object detection
and photometry of the $NB1060$ and Broad Band WIRCam images. 
The magnitudes were computed in apertures 5 pixels (1.5\arcsec)  in diameter.
We used the CFHT-LS public images of the field for the optical $u^{\ast}g'r'i'z'$ bands,
photometrically corrected as explained in the previous section.

The limiting magnitude of the $NB1060$, $J$ and $Ks$ WIRCam 
observations was estimated as follows: we added 200 artificial 
star-like objects per bin of 0.1 mag on the stacked $NB1060$ image
in carefully selected blank regions of the image. We then ran SExtractor
on this image using the same parameters as previously used for object
detection. Counting the number of artificial stars retrieved in
each magnitude bin provided a direct measure of our completeness limit. 
We report in this paper the limiting magnitude at the 50\% completeness limit.

In the optical we re-binned the original CFHT-LS images with 
0.19\arcsec $\, \textnormal{pixel}^{-1}$
images to the 0.3\arcsec $\, \textnormal{pixel}^{-1}$ scale of the WIRCam images. We then ran
Sextractor with the same parameters used with the WIRCam data. We checked
that the photometry before and after re-binning was preserved. 
We then derived the completeness limit at 50\% for the CFHT-LS images using the same methodology as described above 
for the WIRCam images.

In order to estimate the signal to noise ratio ($SNR$) of our candidates
and the $SNR$ corresponding to our 50\% completeness limit
we used the noise image (BACKGROUND\_RMS) produced by SExtractor. 
This image gives for each pixel the local noise $\sigma$. The $SNR$ 
of an object with number of counts $F$ in an aperture of $A$ pixels is given by:
\begin{eqnarray}
SNR = F/\sqrt{A\sigma^2}
\label{eq:snr}
\end{eqnarray}
and the error on the magnitude $m$ by: 
\begin{eqnarray}
\Delta m=1.086/SNR
\label{eq:err_mag}
\end{eqnarray}
Finally, the limiting magnitude reached by the $NB1060$ observations is 
25.2 in apertures 1.5\arcsec\ in diameter at the 50\%
completeness limit. This limit corresponds to a SNR of $\sim$ 4 and, 
converted into a pure emission line flux, to 
$8.3 \times 10^{-18}\,\textnormal{ergs\,s}^{-1}\,\textnormal{cm}^{-2}$. 

The same procedure was used to derive the limiting magnitudes of all
CFHT-LS and WIRCam images. They are reported in table~\ref{tab:mag}.

\begin{table}
\caption{Observational data.} \label{tab:mag}
\centering
\begin{tabular}{c c c c} \hline \hline
Instrument    & Band  & Integration            & Limiting \\ 
              &         &  time (hrs) & magnitude$^{\mathrm{a}}$\\ \hline 
MegaCam   &         $u^\ast$     &     20.7           &  27.9\\ 
MegaCam   &         $g'$     &       25        &  28.1\\
MegaCam   &         $r'$     &       49         &  27.8\\
MegaCam   &         $i'$     &       74         &  27.4\\
MegaCam   &         $z'$     &       55.8         &  26.5\\ \hline
WIRCam    & NB 1st epoch  &           20           &  24.8\\
WIRCam    & NB 2nd epoch  &           20           &  24.9\\
WIRCam    & NB combined   &           40           &  25.2\\
WIRCam    &     $J$     &     6.2     &  25.0$^{\mathrm{b}}$ \\ 
WIRCam    &     H     &       7.7      &  24.7\\ 
WIRCam    &     $Ks$    &    8.9  &  24.7$^{\mathrm{b}}$ \\ \hline
IRAC      &  3.6 $\mu$m   &      0.034           &  22.2 \\
IRAC      &  4.5 $\mu$m   &     0.034           &  21.5 \\ \hline
\end{tabular}
\begin{list}{}{}
\item[$^{\mathrm{a}}$] $4 \sigma$ magnitude limits in apertures 1.5\arcsec\
in diameter for MegaCam and WIRCam. These limits correspond to a 50\% 
completeness level. IRAC limiting magnitudes are 5$\sigma$ in 3.8\arcsec\ 
apertures.
\item[$^{\mathrm{b}}$] In one quarter of the field, for which additional 
NTT/SOFI data were available, magnitude limits of 25.2 in $J$ and 24.8 in $Ks$ 
were achieved.
\end{list}
\end{table}

\section{Sample construction}
\subsection{Initial candidate selection} \label{sec:initial}
We started by matching all catalogs corresponding to  individual images
to the $NB1060$ catalog using a matching radius
between  positions of 0.7\arcsec\ to allow for  astrometric 
errors.

Our initial selection of \La candidates was based on the following 
criteria:
\begin{enumerate}
\item We selected  objects detected in the $NB1060$ image which do not have
counterpart in any of the optical images ($u^\ast$, $g'$, $r'$, $i'$, $z'$),
as it is indeed virtually impossible to detect any flux blueward of the
\La line. 
Negligible amounts of radiation are expected to escape the galaxy and
to be transmitted by the IGM below the $z = 7.7$ Lyman limit at $\sim$ 790 nm,
and therefore to be detected in the $u^\ast$, $g'$ and $r'$ bands.
All the radiation between the \La and Ly$\beta$ lines at $z = 7.7$ is entirely 
reshifted beyond the Gunn-Peterson (GP) trough at $\sim$ 850 nm observed in 
the spectra of high-redshift quasars \citep{Fan2006}, and which 
corresponds to \La absorption by the partially neutral IGM above $z \sim 6$. 
There should therefore be no detectable flux in the $z'$ band.
Moreover, all photons redder than the Lyman limit at $z = 7.7$ are redshifted 
above the Ly$\beta$ and the \La troughs at $z = 6$, with therefore an optical
depth of 5 or more \citep{Fan2006} leading to
very strong color breaks $i' - J \ga$ 5 or more.

\item We required that the $NB1060$ objects detected in the combined image be 
also detected in each of the half $NB1060$ stacks corresponding to each epoch. 
While each half-stack is at lower $SNR$ than
the total combined image used for generating the master $NB1060$ catalog, this
criterion allows one to remove variable (in flux or in position) objects and
reduces considerably the number of low $SNR$ detections. 
\item We required a signal to noise ratio of $\sim$ 5 or higher on the 
combined image, corresponding to a $SNR \ga 3.5$ in half stack images.
\end{enumerate}

Considering the limiting magnitude of our optical and NB images, the selection
criterion 1 above corresponds to: 
\begin{equation}
\begin{array}{lcl}
u^\ast - NB1060 & > & 2.7 \\
g' - NB1060 & > & 2.9 \\
r' - NB1060 & > & 2.6 \\
i' - NB1060 & > & 2.2 \\
z' - NB1060 & > & 1.3
\end{array}
\label{eq:color_indices}
\end{equation}
Taken altogether, the color break between the optical and $NB1060$  filters
is extremely high and covers a wide spectral range. For the CFHT-LS,
the Terapix data center generated deep $\chi^2$ images combining the $g'$, $r'$ 
and $i'$ images. 
Without entering into considerations on wide band / multi-color 
magnitudes, we can infer from equations~\ref{eq:color_indices} that 
the optical dropout selection can be broadly expressed as: 

\begin{eqnarray}
g'r'i' - NB1060 \ga 3
\label{eq:break}
\end{eqnarray}

We note that this color break is significantly stronger than what has been 
usually achieved in other high-$z$ LAEs or LBGs searches, however with a slight
gap in wavelength between the optical red end and the $NB1060$  filter.

After this initial criterion, a careful visual inspection of the candidates
allowed us to remove a few obvious fake candidates in the form of electronic
ghosts or artifacts around bright stars. A couple of objects of dubious
quality in one or some of the images, or with unusual morphology, were 
also removed.

Finally, three initial candidates that were bright in the $Ks$ band image
with $NB1060 - Ks \ga 1.2$
were interpreted as Extremely Red Objects (EROs) (see also
section~\ref{sec:eros}) and were therefore discarded, corresponding de facto
to using an additional color selection criterion for the candidates: 
\begin{eqnarray}
NB1060 - Ks \la 1.2
\label{eq:K}
\end{eqnarray}
From these simple criteria we derived an initial list of 8 objects, none
of them being resolved at the level of the image quality of the $NB1060$
image (0.76\arcsec). 

Finally, we note that none of the objects in our list do have counterparts
in any of the Spitzer/IRAC SWIRE data. If the H$\alpha$ line was $\sim$ 100 
times brighter than the \La line, it would still be too faint to be detected 
alone in the 5.8 $\mu$m IRAC band to the sensitivity achieved by the SWIRE data.

\subsection{The sample}\label{sec:sample}
\begin{figure*}
\centering
\includegraphics[width=16.cm]{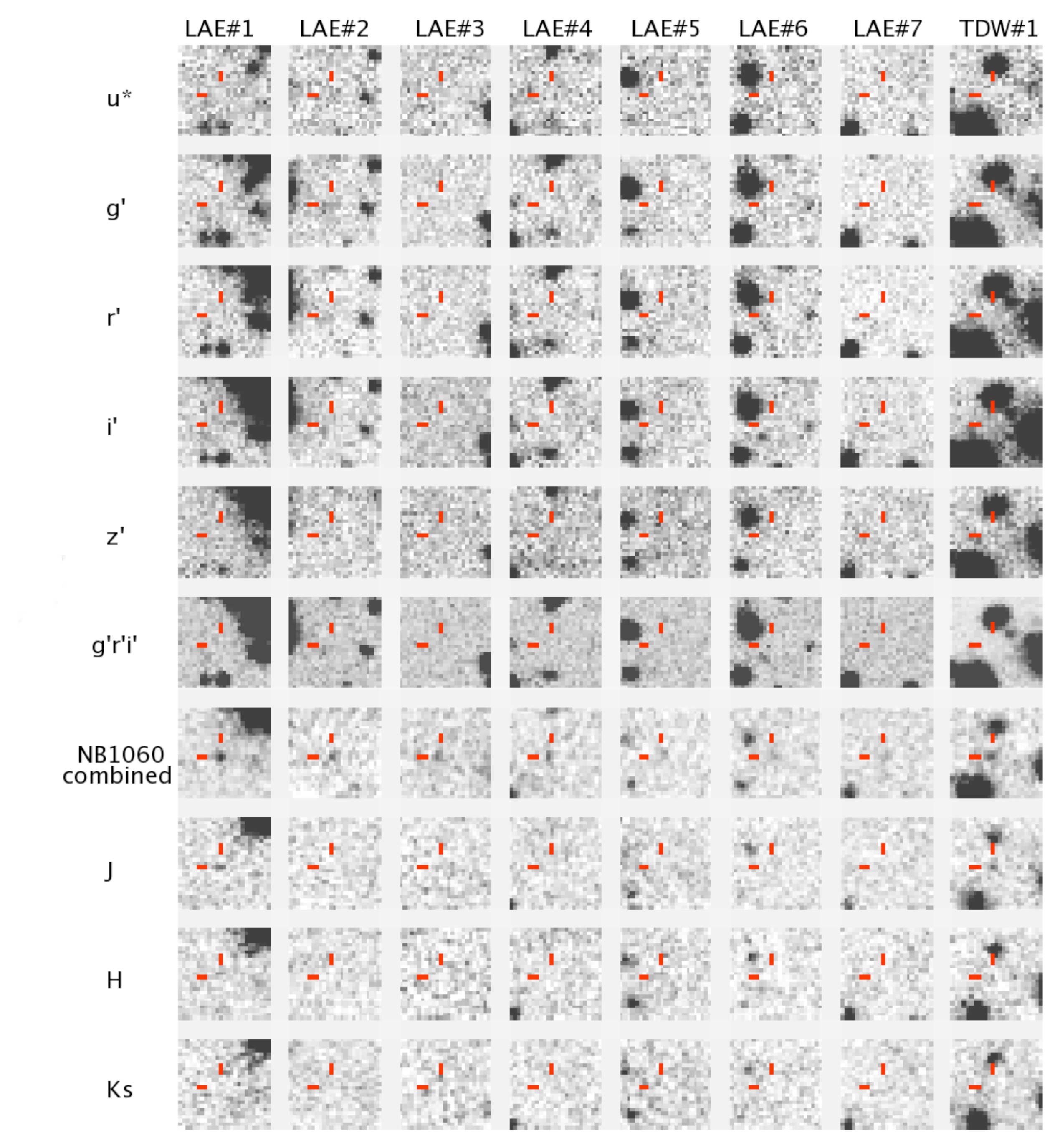}
\caption{Thumbnail images of all candidates in the final sample listed
in table~\ref{tab:cand}. Object TDW\#1 is displayed for reference but
is not part of the LAE sample (see text). Objects Id's and photometric bands
are indicated.}
\label{fig:thumbnails}
\end{figure*}

Our sample consists of 8 objects listed in table~\ref{tab:cand} and 
shown in Fig.~\ref{fig:thumbnails}. 
One has $NB1060 - J \sim$ 0 and therefore 
does {\it not} qualify as an emission line 
object, and is instead identified as a T-dwarf candidate 
(see section~\ref{sec:tdwarfs}).
Only 5 objects have $NB1060 - J <$ 0 and therefore qualify, 
a priori, as emission line objects, while two other 
objects have $NB1060$ fluxes fainter than the detection limit in $J$, and 
therefore cannot be surely identified as line emitting objects.
Therefore, from the six brightest objects of the original sample of eight  selected without using the $J$ magnitudes, five seem to be line emitters. With the same success rate of 5/6 the two faintest objects should therefore also be line emitters and it is therefore reasonable to keep them in the final sample, however flagging them as less secure than the other candidates. 

We also report in table~\ref{tab:cand} the lower limits of the restframe
Equivalent Widths ($EW$) derived from the photometric data, defined as:

\begin{eqnarray}
EW_\mathrm{rest} = \left(\frac{f_{\lambda,NB1060} \times \lambda_{NB1060}^2}{f_{\lambda,J} \times \lambda_J^2} - 0.5\right) \times \frac{\Delta\lambda_{NB1060}}{1 + z}
\label{eq:ew}
\end{eqnarray}

where $f_{\lambda}$ is the observed flux density in 
$\textnormal{ergs\,s}^{-1}\,\textnormal{cm}^{-2}\textnormal{\AA}^{-1}$ 
at the wavelengths of
the $NB1060$ and $J$ filters and $\Delta\lambda_{NB1060}$ is the width of the 
$NB1060$ filter (100 $\AA$). 
It is assumed that the UV continuum is completely extinguished below the Ly$\alpha$ line, and therefore contributes to the $NB1060$ flux, in average, over only half of the filter spectral width. A spectral energy distribution 
$f_{\nu} = \mbox{const.}$ is assumed. Assuming $f_{\lambda} = \mbox{const.}$ 
would lead to $EW$ values approximately twice as large. 
Six out of seven of our LAE photometric candidates are not detected 
in the $J$ band and we therefore use the detection limit in this band,
deriving in turn lower limit $EW$ values.We note that LAE\#6 and LAE\#7 have 
positive $EW$ values despite the fact that their
$NB1060$ magnitudes are fainter than their J magnitudes.

Samples of emission line selected galaxies are nominally defined in terms of 
the limited equivalent width sampled by a particular survey. 
For example, \citet{Taniguchi} present a sample of 9 spectroscopically 
confirmed LAEs at $z = 6.5$ with $EW$ values $EW_\mathrm{obs} > 130\AA$ 
or $EW_\mathrm{rest} > 17\AA$.
Of our sample of candidate $z = 7.7$ LAEs, the faintest certain line emitters 
(e.g. LAE\#5) presents $NB1060 -J < -0.3$ which corresponds to an $EW$ limit 
$EW_\mathrm{obs} > 80\AA$ or $EW_\mathrm{rest} > 9\AA$.
Considering that in all but one case all the $EW$ values are lower limits, 
the lower range of $EW$s sampled by our observations is comparable to that of other studies.
Within the practical limitation of matching the selection criteria of two 
different surveys, the two populations of LAEs revealed by \citet{Taniguchi}
and the current study are therefore approximately equivalent in terms of the $EW$ 
sampled. 

However, the $z = 7.7$ LAEs presented in this paper are selected to be 
$NB1060$ excess sources at a lower significance level than the
\citet{Taniguchi} LAEs, and in addition our sources are not confirmed 
spectroscopically. Therefore, when comparing the LF properties of the $z = 7.7$ 
LAE candidates to confirmed LAE sources at $z = 6.5$ we must include an 
assessment of the unknown sample contamination.

\begin{table*}
\caption{Table of the z$\sim7.7$ LAE and T-dwarf candidates.}
\label{tab:cand}
\centering
\begin{tabular}{c c c c c c c c c c c c} \hline \hline
Id.   & $NB1060$ & Error & SNR ($NB1060$) & $J$  & Error & $SNR (J)$ & $H$  & Error & $SNR (H)$ & $Ks$ & $EW^{\mathrm{a}}$ (\AA)\\ \hline 
LAE\#1 & 24.0 & 0.08 & 14.5 & 24.5    & 0.16  & 6.7  & 24.7 & 0.3 &  4 & $>$24.7 &~13\\
LAE\#2  & 24.3 & 0.17 & 6.5  & $>$25.0 & $--$  & $--$ & $>$24.7 & $--$  & $--$ & $>$24.7 & $>$16\\
LAE\#3 & 24.6 & 0.15 & 7.2  & $>$25.2 & $--$  & $--$ & $>$24.7 & $--$  & $--$ & $>$24.8 &$>$15\\
LAE\#4 & 24.8 & 0.19 & 5.8  & $>$25.2 & $--$  & $--$ & $>$24.7 & $--$  & $--$ & $>$24.8 &$>$11\\
LAE\#5 & 24.9 & 0.2  & 5.5  & $>$25.2 & $--$  & $--$ & $>$24.7 & $--$  & $--$ & $>$24.8 &$>$9\\ \hline
LAE\#6$^{\mathrm{b}}$ & 25.1 & 0.19 & 5.9  & $>$25.0 & $--$  & $--$ & $>$24.7 & $--$  & $--$ & $>$24.7 &$>$5\\ 
LAE\#7$^{\mathrm{b}}$ & 25.1 & 0.22 & 4.9  & $>$25.0 & $--$  & $--$ & $>$24.7 & $--$  & $--$ & $>$24.7 &$>$5\\ \hline
TDW\#1$^{\mathrm{c}}$ & 24.3 & 0.12 & 9.4  & 24.2 & 0.15 & 7.3  & $>$24.7 & $--$  & $--$ & $>$24.7 &-- \\ \hline
\end{tabular}
\begin{list}{}{}
\item[$^{\mathrm{a}}$] Restframe
\item[$^{\mathrm{b}}$] These two objects are not categorically identified as 
line emitting objects, but are still probably LAEs, see text
\item[$^{\mathrm{c}}$] This object is not formally part of
the sample because it is likely a late type T-dwarf, see text
\end{list}
\end{table*}

\subsection{Possible sources of contamination}
Known astronomical objects such as extremely 
red objects or T-dwarfs can potentially satisfy the optical dropout selection 
and therefore possibly contaminate our sample. We examine
various such examples in this section, along with artificial sources of 
contamination.

\begin{figure}
\resizebox{\hsize}{!}{\includegraphics{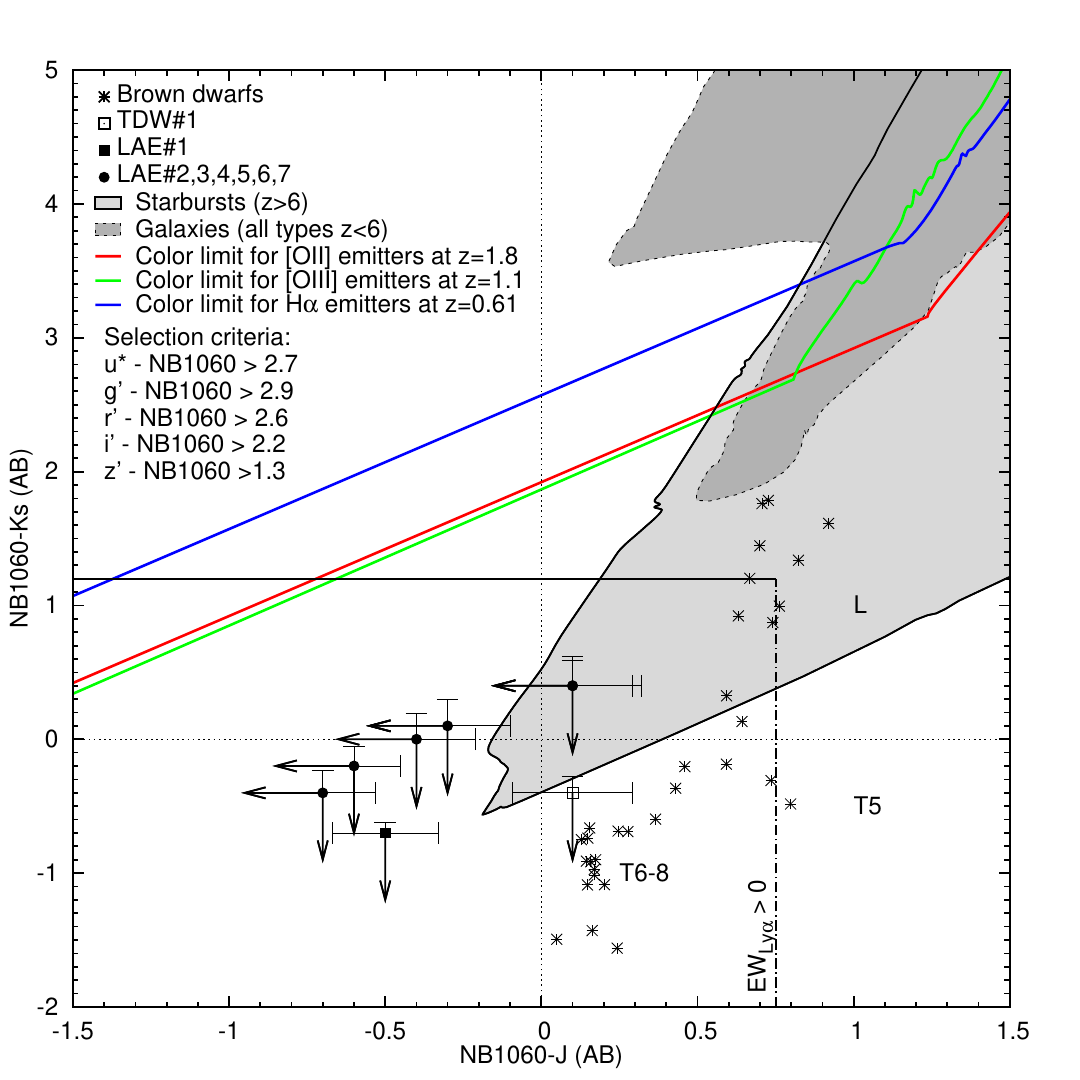}}
\caption{
$NB1060 - J$ and $NB1060 - Ks$ color-color diagram. 
The filled square symbol corresponds to LAE\#1, 
the filled circles to LAE candidates \#2 to \#7, and the open square to the 
T-dwarf candidate. Photometric errors are represented by the error bars
or arrows depending on whether the objects are detected or not in the J and/or
Ks bands. Star symbols correspond to  L and T-dwarfs 
(see section~\ref{sec:tdwarfs}).
The grey areas correspond to the colors of $z > 6$ starbursts and $z < 6$ 
galaxies with arbitrary reddening
values $E(B-V)$ in the range [0--5] that satisfy the color selection criteria
used in this work. The template spectra used to compute the
colors are described in \citet{Ilbert09} and are without emission
lines. 
When adding an emission line to the spectra, the 
points move to the left of the plot parallel to a line of slope 1. The corresponding lines drawn from
the bluest $NB1060 - Ks$  datapoints at redshifts 0.61 (blue), 1.1 (green) and 1.8 (red) are
indicated. Emission line galaxies at these redshifts are located 
above these  lines.}
\label{fig:color}
\end{figure}

In the discussion which follows we will make use of figure~\ref{fig:color} 
which shows the $NB1060 - Ks$ versus $NB1060 - J$ 
colors of the candidates together with other astrophysical objects.

We used the template spectra (without emission lines) described in Ilbert et al. (2009). We then arbitrarily redshifted them in the redshift range [0--8] and we added arbitrary reddening values $E(B-V)$ in the range [0--5], generating in total more than 100,000 spectra. We then applied our selection criteria (equation~\ref{eq:color_indices}), computing for the remaining spectra the $NB1060 - J$ and $NB1060 - Ks$ colors. The envelope of the points corresponding to galaxies at redshifts $< 6$ (respectively $> 6$) is indicated by the dark (resp. light) grey zone in figure~\ref{fig:color}.

\subsubsection{Electronic crosstalk}
The WIRCam detectors suffer from relatively strong electronic cross-talk.
The main effect consists in the appearance of ghost images every 32 
rows or columns around bright and saturated stars. 
Three types of electronic ghosts
have been identified: positive, negative and `edge' ghosts, the latter 
consisting of spot images positive on one edge and negative on the other edge.
The pattern of these electronic ghosts follows the sky on the images since it is
directly associated with (bright) sky objects.
Only the positive ghosts are likely to generate false candidates. They are
however easily recognizable since they are distributed along columns with
a fixed pattern originating from the brightest stars.
Median filtering every other 32 detector rows 
-- corresponding to the number of amplifiers -- allows one to highlight the 
ghosts while removing the real objects. This allows one to easily identify most
of the ghosts present in the images and to ignore them at the source -- this
is what was done and most of the ghosts did not contaminate our initial 
samples of candidates. After careful further investigation, 2 suspicious
objects were discarded and accounted for as electronic ghosts for presenting
several similarities with the pattern of brighter ghosts.

\subsubsection{Persistence}
Another well-known effect of the IR detectors is that pixels illuminated by
bright and / or saturated stars in one image  continue to release electrons
long after the illumination has stopped. This generates fake objects at the
positions once illuminated by  bright stars previously during the night.
These fake sources remain fixed on the detector and therefore do not follow 
the sky during the dithering pattern. In principle they are removed 
by sigma clipping or min-max rejection when the images are stacked, however
faint residuals may remain. Indeed a pattern of faint objects reproducing
the pattern of telescope offsets was observed around the  brightest stars
in the image, generating false candidates which could be easily  removed.
 
\subsubsection{Noise}
Following the approach from \citet{Iye2006}, we have $\sim$ 10$^6$ 1.5\arcsec\
(diameter) circular apertures in our NB data. Since all of our candidates
have SNR $\ga$ 5, and assuming a gaussian
distribution  of the noise, the probability of false 5$\sigma$ events is 
$\sim$ 0.3. This is admittedly a crude analysis which does not take into account
the fact that the noise properties of the stacked and resampled NB image 
deviate from a gaussian distribution. If contamination by noise might possibly 
take place at very low level close to the detection limit, noise certainly 
cannot be a plausible source of contamination for higher SNR objects.

\subsubsection{Transient Objects}

At the flux limit of the survey, distant supernovae can be visible for 
several weeks, and  therefore be a potential source of contamination. \\
We computed the expected number of Type Ia and Type II supernovae that would be visible in our narrow-band WIRCam images, by using the same method presented in \citet{Cuby2007}.
For a limiting magnitude of 24.8 corresponding to one year of data (see 
table~\ref{tab:mag}), we find that $\sim$ 3 supernovae can be present in the 
area covered by our data. While contamination by SNe is therefore 
probable in individual one-year stacks, such objects are automatically removed
from our final list of candidates because of the condition that they be present
in both 1-yr images. Similarly, stacking of data acquired over long time spans
automatically removes point-like slowly moving solar system objects.

\subsubsection{T-dwarfs}\label{sec:tdwarfs}
Using the \cite{Tinney} spectral type vs. absolute magnitude relations
we infer that we could detect T-dwarfs from 300 to 1000 pc, depending on 
spectral type, from the coolest to the warmest. 
Considering the high galactic latitude of our field, this 
extends our sensitivity to T-dwarfs far beyond the Galactic disk scale height. 
Truncating to a Galactic disk scale height of 350 pc applicable to the 
population of T-dwarfs \citep{Ryan}, we derive a sampled volume of
$\sim$ 400 pc$^3$. 
Considering a volume density of T-dwarfs of a few 
10$^{-3}$ pc$^{-3}$, we expect no more than one T-dwarf in our field.
At a couple of hundreds 
of parsecs from the Sun the proper motion of these objects would not be 
detected over a one year timescale.

We  used the public library of L and T-dwarfs spectra compiled by A. Burgasser 
(http://web.mit.edu/ajb/www/tdwarf) to compute the $NB1060 - J$ colors expected
for these objects. Including T-dwarfs as late as T8 (for which NIRC 
spectra are available), the colours satisfy (see figure~\ref{fig:color}) :
%
\begin{eqnarray}
(NB1060 - J)_\mathrm{T-dwarfs} > 0
\label{eq:t_dwarfs}
\end{eqnarray}

One of the brightest candidates detected in the $J$ band has $NB1060 - J$ 
colors satisfying this criterion and is therefore classified as a late type
T-dwarf (TDW\#1). The two objects  LAE\#6 and \#7 have $NB1060 - J$ 
upper limit values also consistent with late type T-dwarfs. However, because
(i) these are only upper limits (ii) we are not expecting many T-dwarfs in our 
data (see section~\ref{sec:tdwarfs}) and (iii) these two objects are likely 
to be line emitters (see section~\ref{sec:sample}), we assume in the 
following that only TDW\#1 is a T6 to T8 
T-dwarf. We show in section~\ref{sec:lbgs} that it could also be, in principle, 
a high-redshift Lyman Break Galaxy, but this does seem less likely.

\subsubsection{Extremely Red Objects} \label{sec:eros}

The extremely red objects (EROs) are usually defined by their $R - K$ 
color, e.g. $R_\mathrm{AB} - K_\mathrm{AB} \ga 3.4$, 
possibly with additional color criteria 
\citep{Cimatti}, and their fluxes increase towards longer wavelengths. 
They are generally identified as either passively evolving, 
old elliptical galaxies, or dusty starbust galaxies \citep{Pierini,Bergstrom}. 
Despite their faintness in the optical bands, the vast majority of the ERO 
population present in our data are detected in the $r'$, $i'$ or $z'$ bands, 
and therefore not selected as LAE candidates. In practice, only 3 objects
which passed our initial selection criteria were identified as EROs based
on their bright red $NB1060 - Ks$ colors, and the three of them are 
spatially resolved. 
Their $Ks$ magnitudes and sizes are given in table~\ref{tab:EROs}.
After removal of these 3 objects from our sample, 
none of the remaining objects are detected in the $Ks$ band.
The $Ks$ band data, although of limited depth, do therefore provide a 
reasonably robust selection tool for discriminating EROs from LAE 
candidates.

\begin{table}
\caption{Extremely Red Objects - $NB1060$, $Ks$ magnitudes and size}\label{tab:EROs}
\centering
\begin{tabular}{c c c c c c} \hline  \hline
Id    & $NB1060^\mathrm{a}$  & Error  & $Ks^\mathrm{a}$     & Error  & Size$^\mathrm{a}$   \\ \hline
ERO\#1  & 25.1    &  0.4   & 22.5  &   0.1  & 3.3      \\
ERO\#2  & 24.2    &  0.3   &  23   & 0.1    & 1.9       \\
ERO\#3 & 24.2    &  0.3   & 22.5  & 0.15    & 1.3     \\ \hline
\end{tabular}
\begin{list}{}{}
\item[$^\mathrm{a}$] MAG\_AUTO as given by Sextractor
\item[$^\mathrm{b}$] FWHM in arcsec in the $Ks$ band
\end{list}
\end{table}

\subsubsection{Low redshift Emitters}\label{sec:lowz}
As shown in Eq.\ref{eq:break} our candidates exhibit a very strong color 
break of about three magnitudes between the optical part of the spectrum 
and 1.06 $\mu$m. Contamination could occur from peculiar 
emission line objects such as strong star forming, low metallicity galaxies
with an emission line redshifted in the NB filter and an underlying continuum 
so faint that it would remain undetected in any of the optical broad band 
filters. The most likely sources of low redshift 
contamination are from H$\alpha$ emitters at $z = 0.61$, [OIII] emitters 
at $z = 1.1$ and [OII] emitters at $z = 1.8$. 
We first make use of figure~\ref{fig:color} 
to evaluate the $EW$ of the emission lines required at these redshifts to contaminate our sample. 
Such emission line galaxies are located above the
colored lines, and for their near-IR colors to be consistent with our data (to the extreme left of the plot)
would require a contribution of the emission lines to the $NB1060$ flux of
several magnitudes, equivalent to restframe equivalent widths of
several hundreds or thousands of Angstroms.
We now discuss each
source of contamination in turn.

\vspace{0.2cm}
\noindent {\it H$\alpha$ emitters at $z = 0.61$}

\noindent
To estimate the number of H$\alpha$ emitters present in our WIRCam image we 
use the H$\alpha$ luminosity function of \citet{Tresse} without reddening 
correction. To our detection limit, and assuming that the NB flux is 
dominated by the H$\alpha$ line flux, we derive that we have $\sim 300$ 
H$\alpha$ emitters in our filter and field of view in the redshift interval 
[0.607--0.623]. The vast majority of these emitters are bright in the optical 
and therefore not selected.
Indeed, using the \citet{Ilbert2005} luminosity functions in the 
restframe UBV filters and in the redshift bin [0.60-0.8] we estimate the 
limiting magnitude providing 300 objects in the comoving volume 
sampled by the H$\alpha$ line through the $NB1060$ filter. 
We obtain magnitudes of 25.8, 26 and 25.6 
corresponding approximately to the $r'i'z'$ filters at redshift 0.6, i.e.
more than 1.5 magnitude below the limiting magnitudes of the CFHT-LS
in the $r'$ and $i'$ filters. 
No normal $z = 0.61$ galaxy spectral energy distribution 
can therefore contaminate our sample.

In the extreme case of a pure line emission spectrum, the sensitivity limits
in the $r'$ and $i'$ bands correspond  to [OII] and [OIII] flux
limits of 
$3\times10^{-18}\,\textnormal{ergs\,s}^{-1}\,\textnormal{cm}^{-2}$ and
$2\times10^{-18}\,\textnormal{ergs\,s}^{-1}\,\textnormal{cm}^{-2}$
respectively. Assuming an intrinsic Balmer decrement of 2.8 for case B
recombination, this would correspond to [OII]/H$\beta$ and [OIII]/H$\beta$
ratios respectively below 1.0 and 0.71. In the sample of galaxies between
redshifts 0.4 and 3 presented in \citet{Maier}, no objects meet these criteria
simultaneously. All objects with low [OIII]/H$\beta$ ratios have strong
[OII]/[OIII] ratios, and either of the two [OII] or [OIII] 
lines should therefore be detected.

In conclusion of this analysis, we argue that contamination by 
$z = 0.61$ galaxies is unlikely to bias significantly our sample.

\vspace{0.2cm}
\noindent {\it [OIII] emitters at $z = 1.1$}

\noindent
At redshifts 1.12, similarly to the case above, the vast majority of the 
[OIII] emitters in the comoving volume 
sampled by the NB filter will be detected
in the optical bands. We use Fig. 13 of 
\citet{Kakazu} to estimate at about 100 the number of [OIII] emitters in our
NB image, assuming no dramatic evolution in the [OIII] LF between redshifts 
0.83 and 1.1 and considering that the line emission dominates the $NB1060$ flux.
Contamination in our sample could only take place from low luminosity, high 
$EW$ [OIII] emitters, unlikely to represent more than a handful of objects. Even
so, for [OII] not to be detected in the $i'$ filter would require a 
[OIII]/[OII] ratio of 4 or higher at the detection limits of the $i'$ and NB 
images. In the data presented in \citet{Maier} at redshifts between 0.4 and 3, 
no more than 5\% of the galaxies have such high values. In total, it is 
therefore unlikely that our sample be contaminated by [OIII] interlopers.

\vspace{0.2cm}
\noindent{\it [OII] emitters at $z = 1.8$}

\noindent
Using the [OII] luminosity function of \citet{Rigopoulou} we estimate the
number of [OII] emitters in our NB image at $\sim$ 300 objects. 
 The [OIII] and H$\alpha$ lines at
redshift 1.84 fall between the atmospheric windows between the $J$ and $H$ bands
for [OIII] and between the $H$ and $Ks$ bands for H$\alpha$, and therefore
do not contribute to the near-IR fluxes in these bands. 
Using the \citet{Kennicutt} relations between the UV continuum and [OII] 
luminosities we derive a rough estimate of the optical magnitudes expected
for [OII] emitters at 1.06 $\mu$m. At the flux limit of our data, we get 
$m_\mathrm{opt}^\mathrm{AB} \approx 25.5$, which 
is 2 to 3 magnitudes brighter than the limiting magnitude of our optical data. 
Even with a large scatter around this value, we expect that 
the vast majority of [OII] emitters should be readily detected in the 
optical bands. Dusty
starbursts may obviously have much fainter optical magnitudes, but such 
objects would fall in the category of Extremely Red Objects, which, 
as explained above, can be discarded from their brightness in the $Ks$ band. 
Very strong,
and unusual [OII] $EW$ would be required for dusty [OII] emitters to be
selected as candidates without being detected in the $Ks$ band, and 
such objects would likely be spatially resolved.

We note that \citet{Taniguchi} suggest that the few line emitters in their 
sample of $z = 6.5$ LAE candidates resisting a definitive identification as LAEs 
could be [OII] emitters. Some, if not all, of these [OII] emitters  
would probably be detected in the $Ks$ band with the 
same detection limit as ours.
We therefore argue that contamination by [OII] emitters in our survey is
likely to be low and unlikely to contaminate a large fraction of our sample.

\subsubsection{High-redshift LBGs} \label{sec:lbgs}
Bright high-redshift Lyman Break Galaxies (LBGs) can be detected in the
$NB1060$ filter through their UV continuum. For this to happen, the redshift
shall be smaller than our target value of 7.7, but high enough for these
objects to escape detection in the optical images, irrespective of the presence
or not of \La emission. See \citet{Cuby2003} for such an example.

To estimate the possible contamination by these bright, high-z UV sources,
we use the \cite{Bouwens} UV LF at z $\sim$ 6 and $\sim$ 7. As a worst
case scenario we consider two redshift ranges: the [6.0--7.0] range 
for which we use the $z = 5.9$ UV LF and the range [7.0--7.7] for which we 
use the $z = 7.3$ UV LF. I band dropouts may fail detection in the z band while
being detected in the $NB1060$ filter.
While the number of objects is $<<$1 in the second
redshift range, it is $\sim$ 3 in the first range. We consider this number
as a significantly overestimated number, for the reason that the simple 
calculation above uses the luminosity function at $z = 5.9$ which applies to 
the lower bound of the redshift range and for which the luminosity is  
brighter. Had we used the $z = 7.3$ LF to match the [6.0--7.0] range, 
we would have found 0.2 objects.

We note that these objects would -- but for their possible detection in the 
$z'$ band --
pass all of our selection criteria, but could be mistaken as late
type T-dwarfs (see section~\ref{sec:tdwarfs}. Interestingly, one of the 
brightest candidates in our sample (TDW\#1), although primarily thought to 
be a T-dwarf, could also be a (bright) LBG.

\subsubsection{Conclusion}

We have analyzed various possible sources of contamination for our sample.
We note that the magnitudes of our candidates
are well distributed, and do not cluster towards the faint end of the
luminosity range probed by our survey. This is in itself a sanity check
demonstrating that we are not sensitive to a sudden increase of the false
alarm rate towards faint fluxes. 
We note that we have made use of very robust selection criteria to select
our candidates, consisting of  very strong color breaks between the optical
and the near-IR ($\sim$ 3 magnitudes), and rejecting EROs from their 
bright fluxes in the $Ks$ band. 
We argue that our selection criteria are comparable to the criteria used 
in other LAE or
LBG studies and we are therefore confident in the reliability of our sample.
We however cannot completely rule out contamination by one or the other
sources identified above, in particular artifacts and / or [OII] emitters.
In the following, we will evaluate the impact of partial contamination of 
our sample, at the level of a couple of objects.

\section{Discussion}
\subsection{Variance}
The variance on the number of objects in our sample is due to Poisson
errors and to fluctuations in the large scale distribution of galaxies.
Various models exist in the literature to account for the effects
of cosmic variance.
\citet{Trenti} have developed a model which is
offered as an on-line `cosmic variance calculator'. Assuming a one-to-one 
correspondence between dark halos and LAEs, we obtain from this model a value 
of $\sim$28\% for the cosmic variance in our data.

This result however strongly depends on the assumptions used for the level of 
completeness and contamination factor in our sample. 
In view of the limited number of objects in our sample and of the 
large comoving volume ($\sim 6.3\times10^{4}$ Mpc$^3$), our results are
probably limited by Poisson noise -- $\sim$38\% for 7 objects --
more than by clustering variance. We note however that variance due to
clumpy re-ionization is ignored and may also contribute to the total
variance.

\subsection{On the Ly$\alpha$ Luminosity function at $z = 7.7$}
From our data we can derive constraints on
the \La luminosity function of $z = 7.7$ LAEs. We first
apply a corrective factor when converting the $NB1060$ magnitudes to
\La fluxes. From the $J$ magnitude of object LAE\#1 we infer that
the \La line  contributes to $\sim$ 70\% of the $NB1060$ flux, a value 
similar to the average value observed for the $z = 6.5$ LAEs of 
\citet{Taniguchi} and which corresponds to an $EW_{\mathrm obs}$ 
of $\sim$ 110 $\AA$ in the observed frame.
We adopt this line to continuum ratio in the following to derive
the \La fluxes from the $NB1060$ magnitudes, adding a 0.1 magnitude rms 
error to account for the dispersion of this ratio between objects, consistent with the dispersion of the EW values of \citet{Taniguchi}.
Clearly, deeper $J$ band imaging or spectroscopy would be required to 
estimate this fraction on a case by case basis for each object. 

To discuss the \La luminosity function derived from our sample we first fit 
a Schechter function $\Phi(L)$ to our 
data, where $\Phi(L)$ is given by:

\begin{equation}
\Phi(L)\mathrm{d}L=\Phi^{*}\left(\frac{L}{L^{*}}\right)^{\alpha}\mathrm{exp}\left(-\frac{L}{L^{*}}\right)\frac{\mathrm{d}L}{L^{*}}
\end{equation}

Considering the scarcity of datapoints in our sample, we do not 
attempt at fitting the 3 parameters of the Schechter function. Following
typical values adopted by \citet{Ouchi} and \citet{Kashikawa} for the 
faint end slope $\alpha$  of the luminosity function, we use $\alpha = -1.5$
and derive $\Phi^{*}$ and $L^{*}$ by a simple $\chi^2$ 
minimization. We note that there may be other functions more representative of
the \La LAE LF at high redshifts. For instance, \citet{Kobayashi} derive
numerical LFs from models of hierarchical galaxy formation and of the escape 
fraction of \La photons from host galaxies.
These LFs differ significantly from a Schechter function. However, for the
sake of comparison with previous work, we keep the analytical Schechter 
formalism.

\begin{figure}
\resizebox{8cm}{!}{\includegraphics{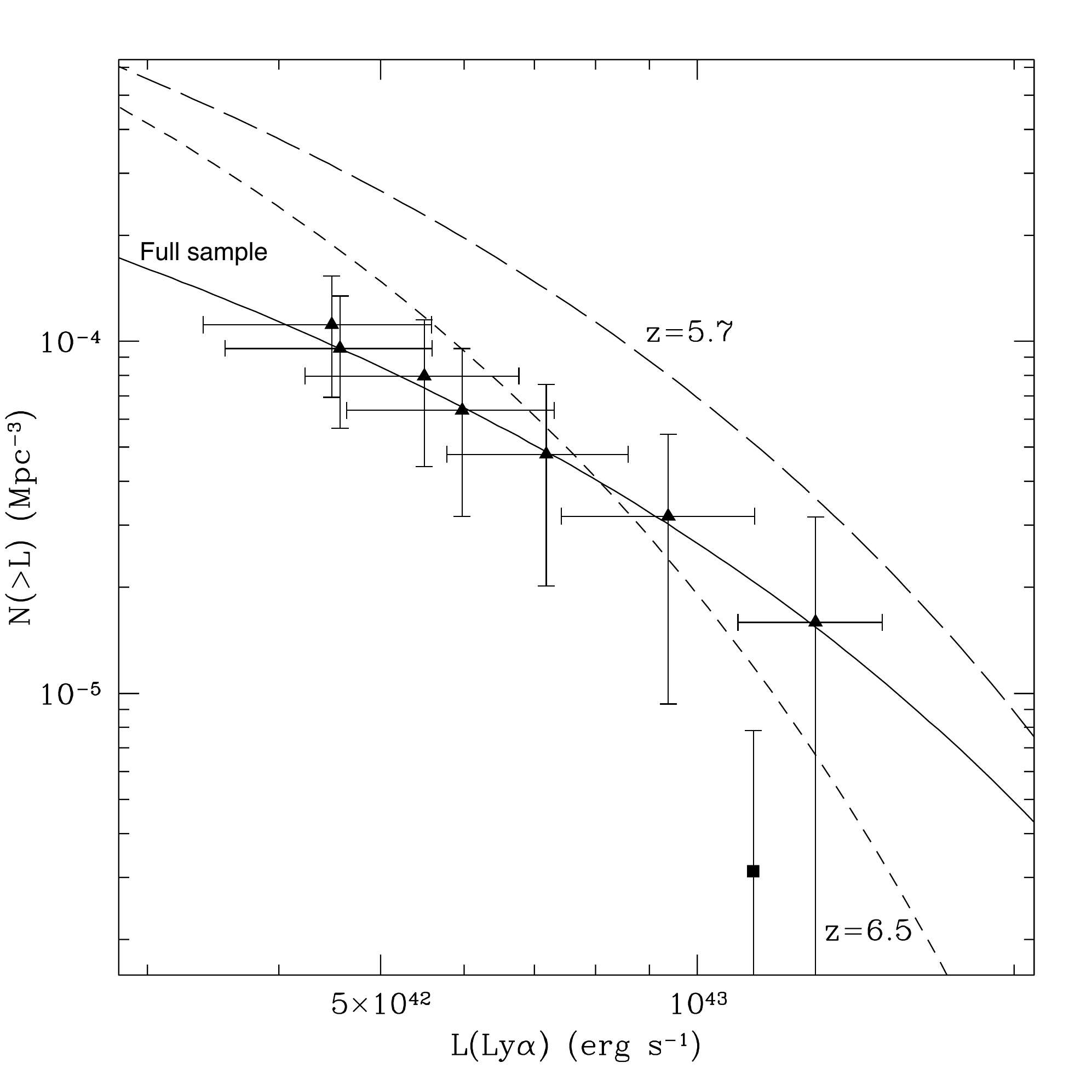}}
\resizebox{8cm}{!}{\includegraphics{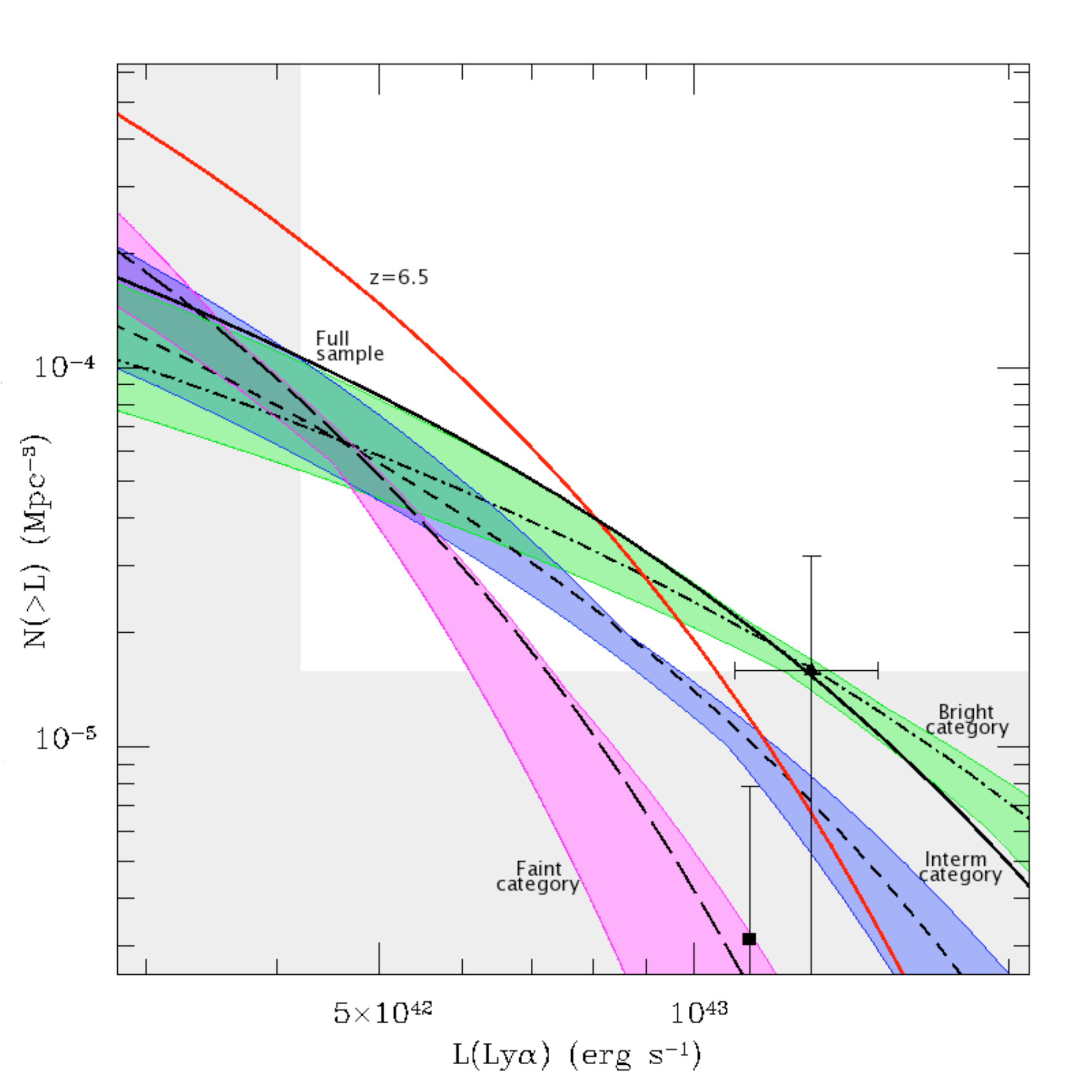}}
\caption{Cumulative Luminosity Functions (LFs) with $\alpha = -1.5$. 
{\it Up}: Luminosity function derived from the {\it full} sample.
The data (triangles) are not corrected for completeness and a factor of 70\% 
has been used to convert the $NB1060$ fluxes into \La fluxes. The plain line is
the best fit Schechter function through the datapoints.
The $z = 5.7$ (\citet{Ouchi}, long dashed line) and the $z = 6.5$
(\citet{Kashikawa}, short dashed line) LFs are also indicated. 
Each point in the sample and its aperture photometric errors, 
augmented by the photometric calibration errors and dispersion of the 
conversion from $NB1060$ magnitudes to \La line fluxes, define the luminosity 
bins and the horizontal error bars. The vertical error bars represent
the Poisson noise associated to the number of candidates.
The square point corresponds to the $z = 6.96$ LAE of by \citet{Iye2006}. 
{\it Bottom}: The shaded color areas represent the range
of LFs in each of the three categories described in the text, and the dashed
lines represent the associated average LFs. {\it Bright} category: green area
and dot-dashed line ; {\it intermediate} category: blue area and short-dashed 
line ; {\it faint} category: magenta area  and long-dashed line. 
Also plotted is the LF at $z = 6.5$ (\citet{Kashikawa}, plain red line) and the 
{\it full} sample $z = 7.7$ LF (this work, plain black line). The square point 
is the $z = 6.96$ LAE of \citet{Iye2006} and the triangle is LAE\#1 (this work).
The shaded grey areas to the bottom and left of the figure represent the
regions where the $z = 7.7$ LF would lie if none of our candidates were real 
LAEs.}
\label{fig:fitlf}
\end{figure}

We initially assume that all of the 7 candidates of the {\it full} sample
are true $z = 7.7$ LAEs, and derive the corresponding LF parameters.
To evaluate the impact of sample contamination on the results, 
we then consider situations where only 4 candidates among the 7 of the sample
are  real. As shown earlier in this paper, despite the robustness of
our sample, we cannot completely rule out contamination by an instrumental 
artifact or a low redshift peculiar object. We therefore conjecture that at 
least 4 objects in our sample must be real $z = 7.7$ LAEs, and
evaluate the impact on the Luminosity Function parameters with such reduced 
samples. To do this, we consider all combinations of 4 objects from among the 
7  in the sample, and compute for each subset the best fit parameters
using a Schechter function. The results of the fits for the full sample and
for the 35 sub-samples are shown on figure~\ref{fig:fitlf}, which shows that
the fits of the 35 sub-samples of 4 objects naturally divide into 3 different
categories:
\begin{itemize}
\item a \textit{bright} category of 20 samples which all contain the
brightest object (LAE\#1)
\item an \textit{intermediate} category of 10 samples which do not 
contain the brightest object but which do contain
the second brightest object (LAE\#2)
\item a \textit{faint} category of 5 samples containing none of the two
brightest objects
\end{itemize}

For each sample we derive the best fit Schechter LF parameters for 
$\alpha = -1.5$ and we derive the average parameters, reported
in table~\ref{tab:fitlf}, for each of the three categories defined above. 
The corresponding average LFs for each category are shown as dashed lines on 
figure~\ref{fig:fitlf}.
We note that the {\it full} sample LF (plain black line) corresponds very 
closely to the LF of the sample of the four brightest objects from the 
{\it bright} category. We also note that by removing three  objects from the 
{\it full} sample while keeping the brightest in the {\it bright} category, 
we artificially enhance the $L^*$ parameter, or alternatively reduce the 
$\Phi^*$ parameter of this category. In consequence the {\it full}
sample LF is probably more representative of the situation where the brightest
object is real than the average LF of the {\it bright} category.

\begin{table}
\caption{Best fit Schechter LF parameters for $\alpha = -1.5$}
\label{tab:fitlf}
\centering
\begin{tabular}{c c c} \hline \hline
Redshift & $\mathrm{log}(L^{*} (\mathrm{erg}\, \mathrm{s}^{-1}))$ & $\mathrm{log}(\Phi^{*}(\mathrm{Mpc}^{-3}))$ \\ \hline
7.7$^{\mathrm{a}}$ & $43.0^{+0.2}_{-0.3}$ & $-3.9^{+0.5}_{-0.3}$  \\
7.7$^{\mathrm{b}}$ & $43.3^{+0.6}_{-0.6}$ & $-4.4^{+1.2}_{-0.7}$  \\  
7.7$^{\mathrm{c}}$ & $42.9^{+0.5}_{-0.7}$ & $-3.8^{+2.1}_{-0.8}$ \\  
7.7$^{\mathrm{d}}$ & $42.5^{+0.5}_{-0.7}$ & $-3.0^{+3.5}_{-1.2}$  \\  
6.5$^{\mathrm{(1)}}$ & $42.6^{+0.12}_{-0.1}$ & $-2.88^{+0.24}_{-0.26}$ \\ 
5.7$^{\mathrm{(2)}}$ & $42.8^{+0.16}_{-0.16}$ & $-3.11^{+0.29}_{-0.31}$  \\\hline
\end{tabular}
\begin{list}{}{}
\item[$^\mathrm{a}$] Full sample of 7 candidates
\item[$^\mathrm{b}$] Mean of the 20 {\it bright} samples of 4 objects among 7
\item[$^\mathrm{c}$] Mean of the 10 {\it intermediate} samples of 4 objects among 7
\item[$^\mathrm{d}$] Mean of the 5 {\it faint} samples of 4 objects among 7, see
text
\item[References.]  (1) \cite{Kashikawa}; (2) \cite{Ouchi}
\end{list}
\end{table}

\begin{figure}
\resizebox{\hsize}{!}{\includegraphics{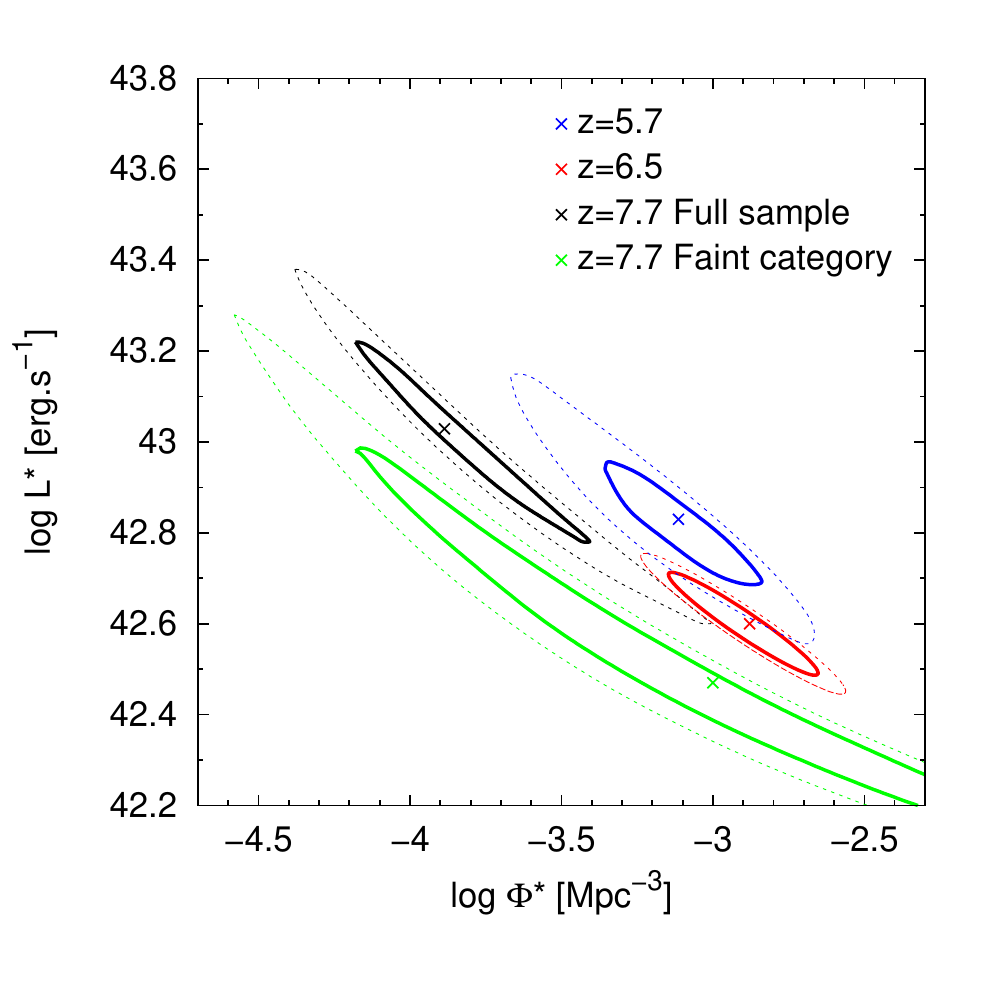}}
\caption{
Error ellipses for the best-fit Schechter parameters $\Phi^{*}$ and $L^{*}$ for $\alpha=-1.5$. The solid and dotted line ellipses are respectively the 68\% and 95\% confidence levels}. Are shown the error contours of the Schechter parameters for the {\it full} sample (black), the {\it faint} category (green), the $z = 6.5$ LF \citep{Kashikawa} (red) and  the $z = 5.7$ LF \citep{Ouchi} (blue).

\label{fig:ellip}
\end{figure}

To further illustrate our results in the light of the LAE LF at lower redshifts,
we plot in figure~\ref{fig:ellip} the error contours for the {\it full} sample 
and the {\it faint} category together with the $z = 6.5$ and $z = 5.7$
LFs from \citet{Kashikawa} and \citet{Ouchi}. The errors for the {\it faint}
category are dominated by the fitting errors for each of the five LFs in the
category, and to account for the dispersion between samples we simply add to  
the fitting errors the deviation between the two most extreme
LFs in the sample. The {\it full} sample LF would indicate an evolution 
between $z = 7.7$ and $z = 6.5$ opposite (in $L^*$ and $\Phi^{*}$) to the 
evolution between $z = 6.5$ and $z = 5.7$ at the 2$\sigma$ confidence level.
Conversely, the {\it faint} category LF would be consistent with this evolution 
and with the $z = 6.96$ datapoint of \citet{Iye2006}. In other words,
for our data to be consistent with other work requires that both of our 
brightest objects are not real LAEs. 

Only spectroscopic confirmation will allow to draw firmer 
conclusions, which will still be based on small numbers and therefore subject
to large uncertainties. Finally, we note that the analysis performed in this
section would remain qualitatively and quantitatively similar had we assumed
that only 3 of our candidates were real instead of 4.

\subsection{Implications}
We used the model from \citet{Kobayashi} to estimate the ionization 
fraction of the IGM at $z = 7.7$ from our constraints on the \La LAE LF.
This model produces LFs for different values of the IGM transmission to 
\La photons ($T_\mathrm{Ly\alpha}^\mathrm {IGM}$) used as a global parameter. 
The \La attenuation by the IGM is a complex process  originating in the 
damping wings of the \La line  and involving the neutral fraction of 
hydrogen $x_\mathrm{HI}$ and dynamical models of the local
IGM infall towards the LAEs \citep{Santos,Dijkstra}.
Within this model the conversion factor from 
$T_\mathrm{Ly\alpha}^\mathrm{IGM}$ to 
$x_\mathrm{HI}$ is therefore highly sensitive to the local IGM density and
velocity, and may not be representative of the {\it average} IGM situation. 
Using this model, to best fit our LFs within the range of observed 
luminosities, we derive $T_\mathrm{Ly\alpha}^\mathrm {IGM}$ values ranging from 
$\sim 0.7$ for the {\it faint} category to 1.0 for the {\it full} sample.
$T_\mathrm{Ly\alpha}^\mathrm {IGM} \sim 0.7$ corresponds to 
$x_\mathrm{HI} \sim 0.3$ in the model of \citet{Santos} for a given redshift 
of the \La line with respect to the systemic velocity of the galaxy. This 
$x_\mathrm{HI}$ value is approximately similar to the one derived from the 
LAE \La LF at $z = 6.5$ \citep{Kobayashi}.

Considering the high level of uncertainty on the $z = 7.7$ LF derived from our
results and on the re-ionization  models, we simply note here, in parallel to
our earlier conclusions on the LF, that if one or both of our brightest 
objects is real, a low fraction of neutral hydrogen is inferred ($\approx 0$), 
in contradiction with earlier reports of an increasing fraction above 
$z \sim 6$. Even if both of the brightest objects are not real 
but if a reasonable fraction of the faint objects is real (our faint
category), the \citet{Kobayashi} model infers a still moderate neutral 
fraction of hydrogen.

\begin{figure}[!ht]
\centering
  \includegraphics[width=7cm]{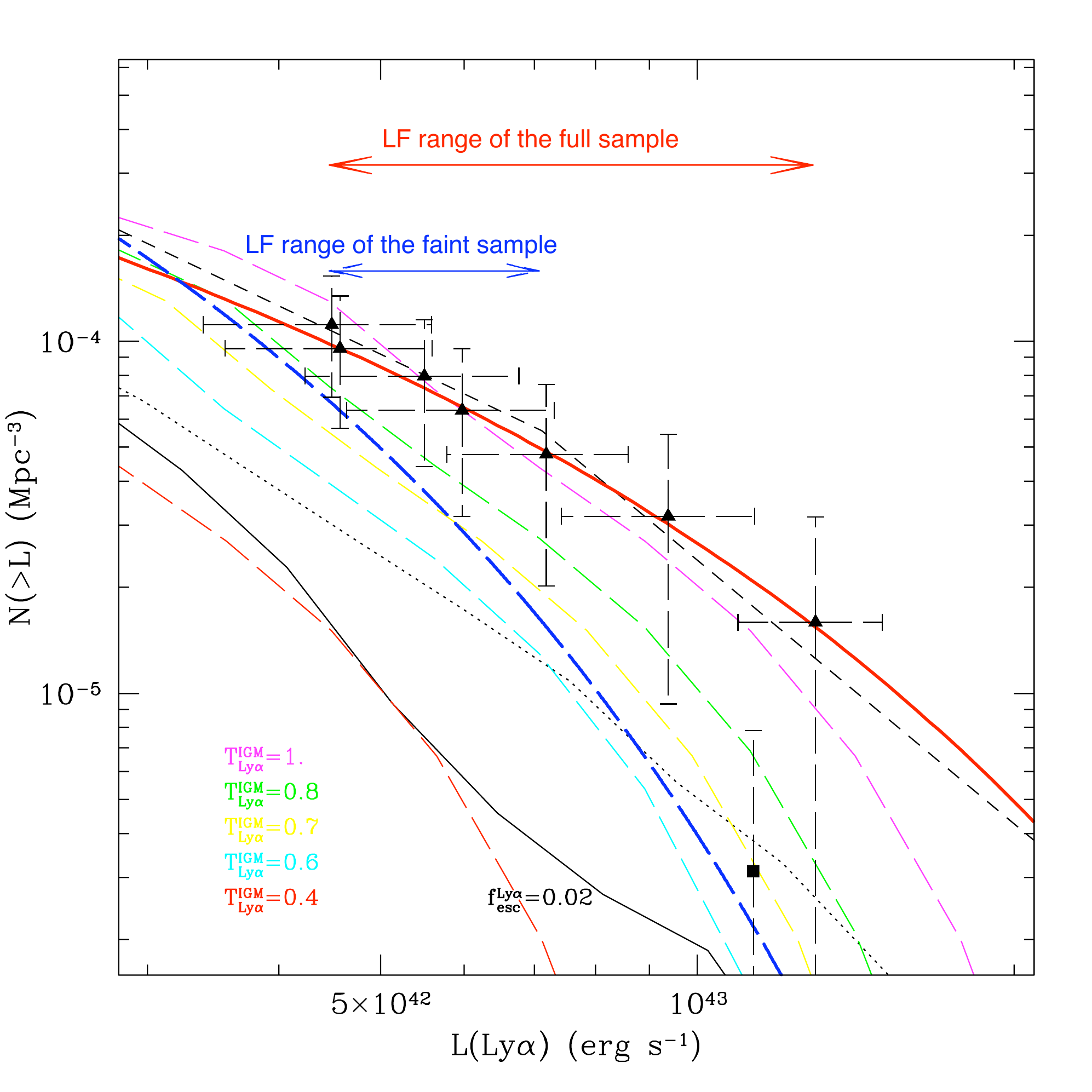}
\caption{
Cumulative Luminosity Functions corresponding to our {\it full} sample (plain
red thick line) and {\it faint} category (long blue dashed thick line).
The range of luminosities sampled by the data for both samples are indicated
by the arrows. Thin lines are z$\sim$8 
LFs from various models: \citet{Baugh} (plain line), \citet{Mao2007}
(dotted line), \citet{Thommes} (short-dashed line),
\citet{Kobayashi} (color long-dashed lines for various IGM \La
transmissions $T_\mathrm{Ly\alpha}^\mathrm {IGM}$) as indicated.
The square point is the $z = 6.96$ LAE of Iye et al. (2006)}
\label{fig:lf_models}
\end{figure}

Finally we note that many other models are available in the literature which 
predict the evolution of the LF of high-z LAEs 
\citep{Baugh,LeDelliou2006,Cole2000,Mao2007,Thommes,Mesinger}, see also
\citet{Nilsson} for a comparison of some of these models. These models
do use various ingredients into the simulations, and discussing our results 
in the light of each of these models is beyond the 
scope of this paper. For the sake of visual comparison, we plot in 
figure~\ref{fig:lf_models} our results corresponding to the extreme 
{\it full} sample and {\it faint} category compared to some of these
models.

\section{Conclusions}

We have searched $z = 7.7$ LAEs in a deep narrow-band image at 1.06 $\mu$m 
obtained with WIRCam at CFHT, totalling 40 hrs of integration time and 
covering 390 arcmin$^2$ and sampling a comoving volume of 
$6.3\times10^{4}$ Mpc$^3$,
down to a limit in \La luminosity of 
$8.3 \times 10^{-18}\,\textnormal{ergs\,s}^{-1}\,\textnormal{cm}^{-2}$. 

Using deep visible data of the field, we have selected LAE candidates presenting
a strong color break of up to 3 magnitudes between the visible data and the 
$NB1060$ filter. We have obtained a sample of seven carefully selected  
candidates.
We have analyzed all sources of contamination that we could think of, and 
argued that contamination is unlikely to affect all of our candidates.

We found that the \La LAE luminosity functions derived from our photometric 
sample, within the limitations of the Schechter formalism and with a fixed
slope parameter $\alpha = -1.5$, would contradict the evolution in luminosity
found by \citet{Kashikawa} between $z = 5.7$ and $z = 6.5$ 
at the 1$\sigma$ to 2$\sigma$ confidence level if only one of our
two brightest objects is real. Only spectroscopic
follow-up of objects in our sample will allow to derive firmer conclusions.

Using models of \La LAE LFs available in the literature, and contingent
upon their reliability, we infer from our results that the fraction of
neutral hydrogen at $z = 7.7$ should be in the range [0.0 -- 0.3].

\begin{acknowledgements}
The authors would like to thank the anonymous referee for constructive comments
which helped us to improve the precision and clarity of the paper ; R. Ellis 
for granting us observing time at the Keck Observatory to attempt spectroscopy 
on an early candidate (which later turned out to be detected on the latest 
CFHTLS release) and J. Richard for performing these Keck observations ; 
J.Mao, A.Lapi, C.Baugh, A.Orsi, M. Kobayashi, E.Thommes, for providing their 
model data and M.Trenti and T.Totani for helpful discussions. We acknowledge 
support from the French Agence Nationale de la Recherche, grant number 
ANR-07-BLAN-0228.

\end{acknowledgements}

\bibliographystyle{aa}
\bibliography{PHibon_astroph2}

\begin{thebibliography}{47}
\expandafter\ifx\csname natexlab\endcsname\relax\def\natexlab#1{#1}\fi

\bibitem[{{Baugh} {et~al.}(2005){Baugh}, {Lacey}, {Frenk}, {Granato}, {Silva},
  {Bressan}, {Benson}, \& {Cole}}]{Baugh}
{Baugh}, C.~M., {Lacey}, C.~G., {Frenk}, C.~S., {et~al.} 2005, \mnras, 356,
  1191

\bibitem[{{Bergstr{\"o}m} \& {Wiklind}(2004)}]{Bergstrom}
{Bergstr{\"o}m}, S. \& {Wiklind}, T. 2004, \aap, 414, 95

\bibitem[{{Bertin} \& {Arnouts}(1996)}]{Bertin}
{Bertin}, E. \& {Arnouts}, S. 1996, \aaps, 117, 393

\bibitem[{{Bouwens} {et~al.}(2009){Bouwens}, {Illingworth}, {Bradley}, {Ford},
  {Franx}, {Zheng}, {Broadhurst}, {Coe}, \& {Jee}}]{Bouwens}
{Bouwens}, R.~J., {Illingworth}, G.~D., {Bradley}, L.~D., {et~al.} 2009, \apj,
  690, 1764

\bibitem[{{Bouwens} {et~al.}(2008){Bouwens}, {Illingworth}, {Franx}, \&
  {Ford}}]{Bouwens2008}
{Bouwens}, R.~J., {Illingworth}, G.~D., {Franx}, M., \& {Ford}, H. 2008, \apj,
  686, 230

\bibitem[{{Cimatti} {et~al.}(2002){Cimatti}, {Daddi}, {Mignoli}, {Pozzetti},
  {Renzini}, {Zamorani}, {Broadhurst}, {Fontana}, {Saracco}, {Poli},
  {Cristiani}, {D'Odorico}, {Giallongo}, {Gilmozzi}, \& {Menci}}]{Cimatti}
{Cimatti}, A., {Daddi}, E., {Mignoli}, M., {et~al.} 2002, \aap, 381, L68

\bibitem[{{Cole} {et~al.}(2000){Cole}, {Lacey}, {Baugh}, \& {Frenk}}]{Cole2000}
{Cole}, S., {Lacey}, C.~G., {Baugh}, C.~M., \& {Frenk}, C.~S. 2000, \mnras,
  319, 168

\bibitem[{{Coupon} {et~al.}(2008){Coupon}, {Ilbert}, {Kilbinger}, {McCracken},
  {Mellier}, {Arnouts}, {Bertin}, {Hudelot}, {Schultheis}, {Le F{\`e}vre}, {Le
  Brun}, {Guzzo}, {Bardelli}, {Zucca}, {Bolzonella}, {Garilli}, {Zamorani}, \&
  {Zanichelli}}]{Coupon}
{Coupon}, J., {Ilbert}, O., {Kilbinger}, M., {et~al.} 2008, ArXiv e-prints

\bibitem[{{Cuby} {et~al.}(2007){Cuby}, {Hibon}, {Lidman}, {Le F{\`e}vre},
  {Gilmozzi}, {Moorwood}, \& {van der Werf}}]{Cuby2007}
{Cuby}, J.-G., {Hibon}, P., {Lidman}, C., {et~al.} 2007, \aap, 461, 911

\bibitem[{{Cuby} {et~al.}(2003){Cuby}, {Le F{\`e}vre}, {McCracken},
  {Cuillandre}, {Magnier}, \& {Meneux}}]{Cuby2003}
{Cuby}, J.-G., {Le F{\`e}vre}, O., {McCracken}, H., {et~al.} 2003, \aap, 405,
  L19

\bibitem[{{Dijkstra} {et~al.}(2007){Dijkstra}, {Lidz}, \& {Wyithe}}]{Dijkstra}
{Dijkstra}, M., {Lidz}, A., \& {Wyithe}, J.~S.~B. 2007, \mnras, 377, 1175

\bibitem[{{Fan} {et~al.}(2006){Fan}, {Carilli}, \& {Keating}}]{Fan2006}
{Fan}, X., {Carilli}, C.~L., \& {Keating}, B. 2006, \araa, 44, 415

\bibitem[{{Ilbert} {et~al.}(2006){Ilbert}, {Arnouts}, {McCracken},
  {Bolzonella}, {Bertin}, {Le F{\`e}vre}, {Mellier}, {Zamorani}, {Pell{\`o}},
  {Iovino}, {Tresse}, {Le Brun}, {Bottini}, {Garilli}, {Maccagni}, {Picat},
  {Scaramella}, {Scodeggio}, {Vettolani}, {Zanichelli}, {Adami}, {Bardelli},
  {Cappi}, {Charlot}, {Ciliegi}, {Contini}, {Cucciati}, {Foucaud}, {Franzetti},
  {Gavignaud}, {Guzzo}, {Marano}, {Marinoni}, {Mazure}, {Meneux}, {Merighi},
  {Paltani}, {Pollo}, {Pozzetti}, {Radovich}, {Zucca}, {Bondi}, {Bongiorno},
  {Busarello}, {de La Torre}, {Gregorini}, {Lamareille}, {Mathez}, {Merluzzi},
  {Ripepi}, {Rizzo}, \& {Vergani}}]{Ilbert}
{Ilbert}, O., {Arnouts}, S., {McCracken}, H.~J., {et~al.} 2006, \aap, 457, 841

\bibitem[{{Ilbert} {et~al.}(2009){Ilbert}, {Capak}, {Salvato}, {Aussel},
  {McCracken}, {Sanders}, {Scoville}, {Kartaltepe}, {Arnouts}, {Floc'h},
  {Mobasher}, {Taniguchi}, {Lamareille}, {Leauthaud}, {Sasaki}, {Thompson},
  {Zamojski}, {Zamorani}, {Bardelli}, {Bolzonella}, {Bongiorno}, {Brusa},
  {Caputi}, {Carollo}, {Contini}, {Cook}, {Coppa}, {Cucciati}, {de la Torre},
  {de Ravel}, {Franzetti}, {Garilli}, {Hasinger}, {Iovino}, {Kampczyk},
  {Kneib}, {Knobel}, {Kovac}, {LeBorgne}, {LeBrun}, {F{\`e}vre}, {Lilly},
  {Looper}, {Maier}, {Mainieri}, {Mellier}, {Mignoli}, {Murayama}, {Pell{\`o}},
  {Peng}, {P{\'e}rez-Montero}, {Renzini}, {Ricciardelli}, {Schiminovich},
  {Scodeggio}, {Shioya}, {Silverman}, {Surace}, {Tanaka}, {Tasca}, {Tresse},
  {Vergani}, \& {Zucca}}]{Ilbert09}
{Ilbert}, O., {Capak}, P., {Salvato}, M., {et~al.} 2009, \apj, 690, 1236

\bibitem[{{Ilbert} {et~al.}(2005){Ilbert}, {Tresse}, {Zucca}, {Bardelli},
  {Arnouts}, {Zamorani}, {Pozzetti}, {Bottini}, {Garilli}, {Le Brun}, {Le
  F{\`e}vre}, {Maccagni}, {Picat}, {Scaramella}, {Scodeggio}, {Vettolani},
  {Zanichelli}, {Adami}, {Arnaboldi}, {Bolzonella}, {Cappi}, {Charlot},
  {Contini}, {Foucaud}, {Franzetti}, {Gavignaud}, {Guzzo}, {Iovino},
  {McCracken}, {Marano}, {Marinoni}, {Mathez}, {Mazure}, {Meneux}, {Merighi},
  {Paltani}, {Pello}, {Pollo}, {Radovich}, {Bondi}, {Bongiorno}, {Busarello},
  {Ciliegi}, {Lamareille}, {Mellier}, {Merluzzi}, {Ripepi}, \&
  {Rizzo}}]{Ilbert2005}
{Ilbert}, O., {Tresse}, L., {Zucca}, E., {et~al.} 2005, \aap, 439, 863

\bibitem[{{Iovino} {et~al.}(2005){Iovino}, {McCracken}, {Garilli}, {Foucaud},
  {Le F{\`e}vre}, {Maccagni}, {Saracco}, {Bardelli}, {Busarello}, {Scodeggio},
  {Zanichelli}, {Paioro}, {Bottini}, {Le Brun}, {Picat}, {Scaramella},
  {Tresse}, {Vettolani}, {Adami}, {Arnaboldi}, {Arnouts}, {Bolzonella},
  {Cappi}, {Charlot}, {Ciliegi}, {Contini}, {Franzetti}, {Gavignaud}, {Guzzo},
  {Ilbert}, {Marano}, {Marinoni}, {Mazure}, {Meneux}, {Merighi}, {Paltani},
  {Pell{\`o}}, {Pollo}, {Pozzetti}, {Radovich}, {Zamorani}, {Zucca}, {Bertin},
  {Bondi}, {Bongiorno}, {Cucciati}, {Gregorini}, {Mathez}, {Mellier},
  {Merluzzi}, {Ripepi}, \& {Rizzo}}]{Iovino}
{Iovino}, A., {McCracken}, H.~J., {Garilli}, B., {et~al.} 2005, \aap, 442, 423

\bibitem[{{Iye} {et~al.}(2006){Iye}, {Ota}, {Kashikawa}, {Furusawa},
  {Hashimoto}, {Hattori}, {Matsuda}, {Morokuma}, {Ouchi}, \&
  {Shimasaku}}]{Iye2006}
{Iye}, M., {Ota}, K., {Kashikawa}, N., {et~al.} 2006, \nat, 443, 186

\bibitem[{{Kakazu} {et~al.}(2007){Kakazu}, {Cowie}, \& {Hu}}]{Kakazu}
{Kakazu}, Y., {Cowie}, L.~L., \& {Hu}, E.~M. 2007, \apj, 668, 853

\bibitem[{{Kashikawa} {et~al.}(2006){Kashikawa}, {Shimasaku}, {Malkan}, {Doi},
  {Matsuda}, {Ouchi}, {Taniguchi}, {Ly}, {Nagao}, {Iye}, {Motohara},
  {Murayama}, {Murozono}, {Nariai}, {Ohta}, {Okamura}, {Sasaki}, {Shioya}, \&
  {Umemura}}]{Kashikawa}
{Kashikawa}, N., {Shimasaku}, K., {Malkan}, M.~A., {et~al.} 2006, \apj, 648, 7

\bibitem[{{Kennicutt}(1998)}]{Kennicutt}
{Kennicutt}, Jr., R.~C. 1998, \araa, 36, 189

\bibitem[{{Kobayashi} {et~al.}(2007){Kobayashi}, {Totani}, \&
  {Nagashima}}]{Kobayashi}
{Kobayashi}, M.~A.~R., {Totani}, T., \& {Nagashima}, M. 2007, \apj, 670, 919

\bibitem[{{Lawrence} {et~al.}(2007){Lawrence}, {Warren}, {Almaini}, {Edge},
  {Hambly}, {Jameson}, {Lucas}, {Casali}, {Adamson}, {Dye}, {Emerson},
  {Foucaud}, {Hewett}, {Hirst}, {Hodgkin}, {Irwin}, {Lodieu}, {McMahon},
  {Simpson}, {Smail}, {Mortlock}, \& {Folger}}]{Lawrence}
{Lawrence}, A., {Warren}, S.~J., {Almaini}, O., {et~al.} 2007, \mnras, 379,
  1599

\bibitem[{{Le Delliou} {et~al.}(2006){Le Delliou}, {Lacey}, {Baugh}, \&
  {Morris}}]{LeDelliou2006}
{Le Delliou}, M., {Lacey}, C.~G., {Baugh}, C.~M., \& {Morris}, S.~L. 2006,
  \mnras, 365, 712

\bibitem[{{Lonsdale} {et~al.}(2003){Lonsdale}, {Smith}, {Rowan-Robinson},
  {Surace}, {Shupe}, {Xu}, {Oliver}, {Padgett}, {Fang}, {Conrow},
  {Franceschini}, {Gautier}, {Griffin}, {Hacking}, {Masci}, {Morrison},
  {O'Linger}, {Owen}, {P{\'e}rez-Fournon}, {Pierre}, {Puetter}, {Stacey},
  {Castro}, {Polletta}, {Farrah}, {Jarrett}, {Frayer}, {Siana}, {Babbedge},
  {Dye}, {Fox}, {Gonzalez-Solares}, {Salaman}, {Berta}, {Condon}, {Dole}, \&
  {Serjeant}}]{Lonsdale2003}
{Lonsdale}, C.~J., {Smith}, H.~E., {Rowan-Robinson}, M., {et~al.} 2003, \pasp,
  115, 897

\bibitem[{{Maier} {et~al.}(2006){Maier}, {Lilly}, {Carollo}, {Meisenheimer},
  {Hippelein}, \& {Stockton}}]{Maier}
{Maier}, C., {Lilly}, S.~J., {Carollo}, C.~M., {et~al.} 2006, \apj, 639, 858

\bibitem[{{Mao} {et~al.}(2007){Mao}, {Lapi}, {Granato}, {de Zotti}, \&
  {Danese}}]{Mao2007}
{Mao}, J., {Lapi}, A., {Granato}, G.~L., {de Zotti}, G., \& {Danese}, L. 2007,
  \apj, 667, 655

\bibitem[{{Marigo} {et~al.}(2008){Marigo}, {Girardi}, {Bressan}, {Groenewegen},
  {Silva}, \& {Granato}}]{Marigo}
{Marigo}, P., {Girardi}, L., {Bressan}, A., {et~al.} 2008, \aap, 482, 883

\bibitem[{{Marmo}(2007)}]{Marmo2007}
{Marmo}, C. 2007, in Astronomical Society of the Pacific Conference Series,
  Vol. 376, Astronomical Data Analysis Software and Systems XVI, ed. R.~A.
  {Shaw}, F.~{Hill}, \& D.~J. {Bell}, 285--+

\bibitem[{{Mesinger} \& {Furlanetto}(2008)}]{Mesinger}
{Mesinger}, A. \& {Furlanetto}, S.~R. 2008, \mnras, 386, 1990

\bibitem[{{Nilsson} {et~al.}(2007){Nilsson}, {Orsi}, {Lacey}, {Baugh}, \&
  {Thommes}}]{Nilsson}
{Nilsson}, K.~K., {Orsi}, A., {Lacey}, C.~G., {Baugh}, C.~M., \& {Thommes}, E.
  2007, \aap, 474, 385

\bibitem[{{Ota} {et~al.}(2008){Ota}, {Kashikawa}, {Malkan}, {Iye}, {Nakajima},
  {Nagao}, {Shimasaku}, \& {Gandhi}}]{Ota}
{Ota}, K., {Kashikawa}, N., {Malkan}, M.~A., {et~al.} 2008, ArXiv e-prints, 804

\bibitem[{{Ouchi} {et~al.}(2008){Ouchi}, {Shimasaku}, {Akiyama}, {Simpson},
  {Saito}, {Ueda}, {Furusawa}, {Sekiguchi}, {Yamada}, {Kodama}, {Kashikawa},
  {Okamura}, {Iye}, {Takata}, {Yoshida}, \& {Yoshida}}]{Ouchi}
{Ouchi}, M., {Shimasaku}, K., {Akiyama}, M., {et~al.} 2008, \apjs, 176, 301

\bibitem[{{Pickles}(1998)}]{Pickles}
{Pickles}, A.~J. 1998, \pasp, 110, 863

\bibitem[{{Pierini} {et~al.}(2004){Pierini}, {Gordon}, {Witt}, \&
  {Madsen}}]{Pierini}
{Pierini}, D., {Gordon}, K.~D., {Witt}, A.~N., \& {Madsen}, G.~J. 2004, \apj,
  617, 1022

\bibitem[{{Richard} {et~al.}(2006){Richard}, {Pell{\'o}}, {Schaerer}, {Le
  Borgne}, \& {Kneib}}]{Richard2006}
{Richard}, J., {Pell{\'o}}, R., {Schaerer}, D., {Le Borgne}, J.-F., \& {Kneib},
  J.-P. 2006, \aap, 456, 861

\bibitem[{{Richard} {et~al.}(2008){Richard}, {Stark}, {Ellis}, {George},
  {Egami}, {Kneib}, \& {Smith}}]{Richard}
{Richard}, J., {Stark}, D.~P., {Ellis}, R.~S., {et~al.} 2008, \apj, 685, 705

\bibitem[{{Rigopoulou} {et~al.}(2005){Rigopoulou}, {Vacca}, {Berta},
  {Franceschini}, \& {Aussel}}]{Rigopoulou}
{Rigopoulou}, D., {Vacca}, W.~D., {Berta}, S., {Franceschini}, A., \& {Aussel},
  H. 2005, \aap, 440, 61

\bibitem[{{Ryan} {et~al.}(2005){Ryan}, {Hathi}, {Cohen}, \& {Windhorst}}]{Ryan}
{Ryan}, Jr., R.~E., {Hathi}, N.~P., {Cohen}, S.~H., \& {Windhorst}, R.~A. 2005,
  \apjl, 631, L159

\bibitem[{{Santos}(2004)}]{Santos}
{Santos}, M.~R. 2004, \mnras, 349, 1137

\bibitem[{{Stark} {et~al.}(2007){Stark}, {Ellis}, {Richard}, {Kneib}, {Smith},
  \& {Santos}}]{Stark}
{Stark}, D.~P., {Ellis}, R.~S., {Richard}, J., {et~al.} 2007, \apj, 663, 10

\bibitem[{{Taniguchi} {et~al.}(2005){Taniguchi}, {Ajiki}, {Nagao}, {Shioya},
  {Murayama}, {Kashikawa}, {Kodaira}, {Kaifu}, {Ando}, {Karoji}, {Akiyama},
  {Aoki}, {Doi}, {Fujita}, {Furusawa}, {Hayashino}, {Iwamuro}, {Iye},
  {Kobayashi}, {Kodama}, {Komiyama}, {Matsuda}, {Miyazaki}, {Mizumoto},
  {Morokuma}, {Motohara}, {Nariai}, {Ohta}, {Ohyama}, {Okamura}, {Ouchi},
  {Sasaki}, {Sato}, {Sekiguchi}, {Shimasaku}, {Tamura}, {Umemura}, {Yamada},
  {Yasuda}, \& {Yoshida}}]{Taniguchi}
{Taniguchi}, Y., {Ajiki}, M., {Nagao}, T., {et~al.} 2005, \pasj, 57, 165

\bibitem[{{Thommes} \& {Meisenheimer}(2005)}]{Thommes}
{Thommes}, E. \& {Meisenheimer}, K. 2005, \aap, 430, 877

\bibitem[{{Tinney} {et~al.}(2003){Tinney}, {Burgasser}, \&
  {Kirkpatrick}}]{Tinney}
{Tinney}, C.~G., {Burgasser}, A.~J., \& {Kirkpatrick}, J.~D. 2003, \aj, 126,
  975

\bibitem[{{Trenti} \& {Stiavelli}(2008)}]{Trenti}
{Trenti}, M. \& {Stiavelli}, M. 2008, \apj, 676, 767

\bibitem[{{Tresse} {et~al.}(2002){Tresse}, {Maddox}, {Le F{\`e}vre}, \&
  {Cuby}}]{Tresse}
{Tresse}, L., {Maddox}, S.~J., {Le F{\`e}vre}, O., \& {Cuby}, J.-G. 2002,
  \mnras, 337, 369

\bibitem[{{Willis} \& {Courbin}(2005)}]{Willis}
{Willis}, J.~P. \& {Courbin}, F. 2005, \mnras, 357, 1348

\bibitem[{{Willis} {et~al.}(2008){Willis}, {Courbin}, {Kneib}, \&
  {Minniti}}]{Willis2008}
{Willis}, J.~P., {Courbin}, F., {Kneib}, J.-P., \& {Minniti}, D. 2008, \mnras,
  384, 1039

\end{thebibliography}

\end{document}